\documentclass[aps,prr,reprint,superscriptaddress]{revtex4-1}
\usepackage{amsmath}
\usepackage{graphicx}
\usepackage{amsfonts}
\usepackage{color}
\usepackage{upgreek}
\usepackage{bm}
\usepackage{lipsum}
\usepackage{comment}
\usepackage{enumerate}
\usepackage{dsfont}

\usepackage[table]{xcolor}
\usepackage{textcomp}
\usepackage{amsmath}
\usepackage{amssymb}
\usepackage{graphicx} 
\usepackage{pgf}
\usepackage{epstopdf}
\usepackage{braket}
\usepackage{mathtools}
\usepackage{times}

\usepackage[retainorgcmds]{IEEEtrantools}
\newcommand{\bea}{\begin{eqnarray}}
\newcommand{\eea}{\end{eqnarray}}

\newcommand{\ii}{\mathrm{i}}

\newcommand{\Tr}{\mathrm{Tr}}

\newcommand*{\defeq}{\mathrel{\vcenter{\baselineskip0.5ex\lineskiplimit0pt\hbox{\scriptsize.}\hbox{\scriptsize.}}}=}

\usepackage[colorlinks,linkcolor=blue,urlcolor=blue,citecolor=blue]{hyperref}

\begin{document}

\title{Effective-Hamiltonian theory: An approximation to the equilibrium state of open quantum systems}

\author{Nicholas Anto-Sztrikacs}
\email{nicholas.antosztrikacs@mail.utoronto.ca}
\thanks{Equal contribution}
\affiliation{Department of Physics and Centre for Quantum Information and Quantum Control, University of Toronto, 60 Saint George St., Toronto, Ontario, M5S 1A7, Canada}

\author{Brett Min}
\email{brett.min@mail.utoronto.ca}
\thanks{Equal contribution}

\affiliation{Department of Physics and Centre for Quantum Information and Quantum Control, University of Toronto, 60 Saint George St., Toronto, Ontario, M5S 1A7, Canada}

\author{Marlon Brenes}
\email{marlon.brenes@utoronto.ca}
\affiliation{Department of Physics and Centre for Quantum Information and Quantum Control, University of Toronto, 60 Saint George St., Toronto, Ontario, M5S 1A7, Canada}

\author{Dvira Segal}
\email{dvira.segal@utoronto.ca}
\affiliation{Department of Chemistry
University of Toronto, 80 Saint George St., Toronto, Ontario, M5S 3H6, Canada}

\affiliation{Department of Physics and Centre for Quantum Information and Quantum Control, University of Toronto, 60 Saint George St., Toronto, Ontario, M5S 1A7, Canada}

\begin{abstract}
We extend and benchmark the recently-developed Effective-Hamiltonian  (EFFH) method [PRX Quantum {\bf 4}, 020307 (2023)] as an approximation to the equilibrium state (``mean-force Gibbs state") of a quantum system at strong coupling to a thermal bath. The EFFH method is an approximate framework. Through a combination of the reaction-coordinate mapping, a polaron transformation and a controlled truncation, it imprints the system-bath coupling parameters into the system's Hamiltonian.
First, we develop a {\it variational} EFFH technique. In this method, system's parameters are renormalized by both the system-bath coupling parameters (as in the original EFFH approach) and the bath's temperature.
Second, adopting the generalized spin-boson model, we benchmark the equilibrium state from the EFFH treatment against numerically-exact simulations and demonstrate a good agreement for both polarization and coherences using the Brownian spectral function.
Third, we  contrast the (normal and variational) EFFH approach with the familiar (normal and variational) polaron treatment. We show that the two methods predict a similar structure for the equilibrium state, albeit the EFFH approach offers the advantage of simpler calculations and closed-form analytical results.
Altogether, we argue that for temperatures comparable to the system's frequencies, the EFFH methodology provides a good approximation for the mean-force Gibbs state in the full range of system-bath coupling, from ultraweak to ultrastrong.
\end{abstract}
\maketitle

\date{\today}
\section{Introduction}

As an emerging field, strong-coupling thermodynamics aims to expand conventional thermodynamics to account for the impact of the 
interaction energy between a quantum system and its surroundings on dynamical and thermodynamical processes \cite{BookQT}.
Strong coupling effects may play a crucial role in determining the behavior of various quantum thermal machines such as heat diodes \cite{diodeSC0,diodeSC1,diodeSC2,diodeSC3}, thermal transistors \cite{transSC1}, engines \cite{Brandes,engineSC,Marti:2018Engine} and refrigerators \cite{RefSC1,RefSC2,RefSC3};
references here are only examples of a rich literature.
Similarly, in complex structures such as light-harvesting systems, the interaction between electronic excitations and the protein complex reveals distinct dynamics compared to the weak coupling regime \cite{lphotoSC}.

However, prior to delving into the exploration of the nonequilibrium functionality of quantum thermal machines, 
we encounter a fundamental question concerning the equilibrium (long-time) state of a stationary system when it is coupled to a heat bath. This equilibrium state, referred to as the mean-force Gibbs state (MFGS)
represents the reduced state of the overall system when traced over environmental degrees of freedom, see Refs.~\cite{Hu:2012Eq,Lutz:2011Eq,JanetPRL,Keeling2022,JanetQC23} for some recent works.
Classically, the Gibbs state is commonly accepted as the equilibrium state of a system interacting with a thermal environment. 
However, when the interaction between the system and the heat bath cannot be neglected, the state of the system is going to deviate 
from this conventional state~\cite{JanetRev}. 

Various methods have been developed to examine the equilibration dynamics and steady state behavior of open quantum systems, going beyond the weak coupling assumption. These methods can be broadly classified into two 
categories: numerically-exact methods; such as the Hierarchical Equations of Motion (HEOM)~\cite{TanimuraHEOM,Kawai2020}, path integral implementations~\cite{TEMPO,Keeling2022}, Monte Carlo methods~\cite{Guy}, generalizations of trace estimators and Krylov subspace methods~\cite{Cheng2022}, and approximate-perturbative techniques~\cite{JanetPRL,Hu:2012Eq,Lutz:2011Eq,JanetRev,JanetQC23,Thoss:2009,Ahsan14,Brumer20,Silbey,ReichS,Cao1,Cao2,Cao3,LatunePRE,LatuneMFGS22,Valkunas20,Becker:2022PRL,Thingna:2012JCP,Thingna:2014JCP,Tupkary:2023PRA,Lee:2022PRE,Correa:2023Shift,Becker:2021PRE,HanggiRev,Miller,Zhang_2022,Giacomo} 
that are commonly developed based on the quantum master equation (QME) formalism~\cite{Nitzan,Breuer}. 

In a recent study~\cite{PRXRC}, we introduced a reaction-coordinate polaron-transform  framework that enables the  
{\it analytical and numerical} study of strong-coupling thermodynamics problems.  
By employing two exact transformations of the Hamiltonian followed by a controlled truncation, this  approach  
 generates an Effective Hamiltonian (EFFH). This new Hamiltonian resembles the original Hamiltonian, with the dimensionality of the Hilbert space being {\it identical} to the starting one, 
but with dressed parameters that depend on the system-bath coupling parameters. 

The essence of the EFFH approach is that 
strong-coupling effects are directly embedded in the system's effective Hamiltonian, which after the transformations and truncation can become weakly coupled to the (residual) surroundings, thus allowing applications of 
weak-coupling techniques to study the resulting dynamics and steady state. 
Altogether, as demonstrated in Ref.~\cite{PRXRC},
the EFFH method allows a direct and easy analysis of strong-coupling effects. It also provides a route for performing 
numerical simulations that include strong-coupling effects nonperturbatively---at the cost of a weak-coupling treatment.

The EFFH method was exercised in Ref.~\cite{PRXRC} on several canonical problems, including models for quantum thermalization and
charge, spin, and energy transport at the nanoscale.
Specifically, it was proposed in Ref.~\cite{PRXRC} that the effective Hamiltonian of the system can be regarded as the Hamiltonian of mean force (HMF)~\cite{HanggiRev,Miller}.
This supposition was supported by demonstrating that a Gibbs state constructed from the system's EFFH was in fact the 
{\it exact} equilibrium state for the spin-boson model
at both the ultraweak and ultrastrong coupling limits~\cite{JanetPRL}.
Furthermore, in the intermediate system-bath coupling regime 
the equilibrium state constructed by the EFFH method agreed qualitatively with the state achieved using the numerical reaction coordinate simulation method.

The analysis in Ref.~\cite{PRXRC} was yet limited in several ways:
First, the equilibrium state of the EFFH was not tested against independent tools, specifically numerically-exact approaches. Second, while the generated EFFH of the system depends on the system-bath coupling parameters, it does not depend on the bath's temperature, contrary to other approaches, specifically the polaron technique \cite{Silbey,Cao1,Cao2,Cao3}. Furthering the development of the EFFH mapping to cover broader regimes of parameters is necessary.
Third, the polaron  transform \cite{Cao1,Cao2,Cao3,Legget} is an established tool for calculating the MFGS of open quantum systems. A relationship between the EFFH method and the standard-popular polaron-frame technique should be
established.

The objective of this study is to develop the Effective-Hamiltonian theory and test its viability as a tool for describing the MFGS of open quantum systems, encompassing the full spectrum of system-bath coupling; the ultraweak, intermediate and strong regimes. We contribute the following:
\begin{enumerate}[(i)]
    \item{Benchmarking: Using the spin-boson model as a case study, we benchmark the EFFH predictions against other methods.
    Particularly, we show that for a Brownian spectral function EFFH results well agree with numerically-exact simulations at arbitrary coupling, a striking result given  that the EFFH state is constructed effortlessly.}
    \item{Extending range of applicability:
    We extend the Effective Hamiltonian method by introducing a variational optimization to the mapping. 
    As a result, the renormalized parameters of the system become dependent on both the spectral properties of the bath and its temperature. Importantly, the variational effective Hamiltonian (var-EFFH) approach 
    does not require the strict separation of energy scales between a small system splittings ($\Delta$) and a large bath characteristic frequency $\Omega$, so long as the coupling energy is weak-to-intermediate.}
    \item{Contrasting: We study in details the relationship between the MFGS as constructed by the EFFH method and the polaron-transformed approach, particularly, the variational flavors of both tools. We show that both the EFFH method and the polaron approach expose analogous physical effects, such as dressing of the system's parameters. However, the EFFH framework offers a clear advantage with the simplicity of its derivation and it catering closed-form expressions for the MFGS.}
\end{enumerate}

Beyond the spin-boson model, we further reinforce the applicability of the EFFH framework by applying it onto a fully harmonic model.

The paper is organized as follows. In Sec.~\ref{sec:Q} we introduce the research question and the principles of the ``normal" EFFH method. We further provide benchmarking against numerically-exact simulations.
In Sec.~\ref{sec:vEFFH} we introduce the principles of the variational EFFH method.
A comparative analysis between the EFFH technique and the polaron-transformed approach is provided in Sec.~\ref{sec:polaron}.
Analytical results for the MFGS and the spin expectation values are given in Sec.~\ref{sec:anal},
further presenting extensive numerical simulations in Sec.~\ref{sec:simul}.
We conclude in Sec.~\ref{sec:Summ}.

\section{Research question and the EFFH method}
\label{sec:Q}

We start by presenting the MFGS in general terms. We then provide an overview of the EFFH technique of Ref.~\cite{PRXRC} and introduce the model we use throughout.

\subsection{The MFGS}
We focus on open quantum systems described by a system's Hamiltonian $\hat{H}_S$ coupled to a  reservoir including a collection of harmonic oscillators. The total Hamiltonian comprising the system, the reservoir and their interaction is given by 
($\hbar\equiv1$)
\bea
\label{eq:generic}
    \hat{H} = \hat{H}_S + \sum_k \nu_k \left( \hat{c}_k^{\dagger} + \frac{t_k}{\nu_k}\hat{S} \right) \left( \hat{c}_k + \frac{t_k}{\nu_k} \hat{S} \right).
\label{eq:h_total_model}
\eea
Here, $\hat{S}$ is an operator defined over the system's degrees of freedom that couples to the reservoir. The bosonic creation (annihilation) operators of the bath are $\hat{c}_k^{\dagger}$ ($\hat{c}_k$) with frequency $\nu_k$ for the $k$-th harmonic mode. The coupling between the system and the reservoir is captured by a spectral density function, $J(\omega) = \sum_k |t_k|^2 \delta(\omega - \nu_k)$. 

It is postulated that the appropriate equilibrium state of the {\it total} Hamiltonian is a Gibbs state at an inverse temperature $\beta=1/T$, given by $\hat{\rho} = e^{-\beta\hat{H}}/\Tr \left(e^{-\beta\hat{H}}\right)$. Here,  we set $k_B\equiv 1$ and ${\rm Tr}$ indicates a full trace over all degrees of freedom. By taking a partial trace over the bath (${\rm Tr_B}$), the state of the {\it system} at equilibrium is given by 
\bea
    \hat{\rho}_{S} = \frac{\Tr_B \left(e^{-\beta\hat{H}}\right)}{\Tr \left(e^{-\beta\hat{H}}\right)}.
    \label{eq:MFGS}
\eea
We refer to this state as the MFGS.
One wonders however whether the state of the system $\hat{\rho}_S$ can be expressed {\it solo} in terms of system's operators as
\bea 
\hat{\rho}_S= \frac{e^{-\beta \hat H_S^*}}{{\rm Tr}_S\left(e^{-\beta \hat H_S^*}\right)}.
\eea
Here, $H_S^*$, an operator of the system, is referred to as the Hamiltonian of mean force and it is typically defined based on this equation.
In this work, we test whether the effective Hamiltonian of the system as constructed in Ref.~\cite{PRXRC}  can serve as $\hat H_S^*$, to offer a good approximation for the MFGS.

We now discuss the different coupling regimes using $\lambda$ to represent the system-bath coupling energy. 
For concreteness, we consider the bath to be characterized by a Brownian spectral density 
function,
\bea
J(\omega) = \frac{4\gamma \Omega^2 \lambda^2 \omega}{(\omega^2 - \Omega^2)^2 + (2\pi\gamma\Omega\omega)^2}.
\label{eq:Brownian}
\eea
It scales with $\lambda^2$ and
it is peaked around $\Omega$ with a width parameter $\gamma$.

First, in the limit of asymptotically-weak coupling, we neglect the interaction energy in the Hamiltonian of Eq.~\eqref{eq:MFGS}.
We then find that the state of the system reduces to the standard Gibbs state of classical physics, 
\bea
    \lim_{\lambda \xrightarrow{} 0}\hat{\rho}_S = \frac{e^{-\beta \hat{H}_S}}{\Tr\left(e^{-\beta \hat{H}_S}\right)}.
    \label{eq:Gibbs_state}
\eea
On the opposite limit, there exists a general expression for the MFGS in the ultrastrong coupling limit~\cite{JanetPRL},
\bea
    \lim_{\lambda \to \infty} \hat{\rho}_S = \frac{e^{-\beta \sum_n \hat{P}_n \hat{H}_S \hat{P}_n}}{\Tr\left(e^{-\beta \sum_n \hat{P}_n \hat{H}_S \hat{P}_n}\right)}, 
    \label{eq:USGibbs_state}
\eea
where now the system thermalizes to a state that is generally non-diagonal in the system's Hamiltonian. Rather, the MFGS is obtained by projecting the system's Hamiltonian onto the eigenspace of the coupling operator $\hat{S}$. 
In Eq.~\eqref{eq:USGibbs_state}, the projectors $\hat{P}_n = \ket{s_n}\bra{s_n}$ are generated from $\ket{s_n}$, the eigenvectors of $\hat{S}$.

Besides these two exact results, there are limited tools capable to {\it analytically} describe the equilibrium state of a general open quantum system. Notably, there are  perturbative expansions developed either from the weak coupling limit~\cite{LatunePRE,LatuneMFGS22,AntonT22} 
or the ultrastrong regime~\cite{AntonUSQME22}.
We next summarize the general idea behind the effective Hamiltonian mapping introduced in Ref.~\cite{PRXRC}, which can tackle thermalization at arbitrary coupling while providing analytical insights.
%=============================

\subsection{Principles of the EFFH method}
\label{subsec:EFFH}

An effective Hamiltonian may be obtained by applying a general transformation to the total Hamiltonian in Eq.~\eqref{eq:h_total_model} followed by subsequent approximations.
The transformation is defined such that the coupling of the system to the reservoir is weakened in the new basis, while the effects of strong coupling become embedded in the system's degrees of freedom. In mathematical language, we apply some operation $\hat{O}$ which may comprise a sequence of transformations and truncations. 
%
%Specific to the EFFH method of Ref. \citenum{PRXRC}, 
The operation $\hat{O}$ is defined such that after its application, the dimension of the system's Hamiltonian is preserved
%the structure of the system's Hamiltonian is preserved, 
but the system-reservoir coupling energy is reduced. 

In a general language, the effective Hamiltonian takes the following form \cite{PRXRC} [compare to Eq.~\eqref{eq:h_total_model}],
\bea
    \hat{H}^{\rm{eff}} &=& \hat{O}\hat{H}\hat{O}^{\dagger}
     \nonumber\\
    &=& \hat{{H}}^{\rm{eff}}_S + \sum_k \omega_k \left( \hat{b}_k^{\dagger} - \frac{2\lambda f_k}{\Omega\omega_k}\hat{{S}} \right) \left( \hat{b}_k - \frac{2\lambda f_k}{\Omega\omega_k} \hat{{S}} \right),
    \nonumber\\
    \label{eq:Heff}
\eea
where  individual terms may be affected from the transformation, but the overall structure is preserved. 
Here, $\omega_k$ are the frequencies of the modes comprising the bosonic bath. The operators $\hat{b}^{\dagger}_k$ and $\hat{b}_k$ are linear combinations of the original harmonic modes.
The coupling energies to this modified bath are given by $f_k$, coupled to the system through the {\it same} system's operator
$\hat{{S}}$. 
The parameters $\lambda$ and $\Omega$ are calculated from the original spectral density function as 
\begin{align}
\label{eq:lambdaOmega}
\lambda^2 = \frac{1}{\Omega} \int_0^\infty {\rm{d}}\omega \; \omega J(\omega), \quad \Omega^2 = \frac{\int_0^\infty {\rm{d}}\omega \; \omega^3 J(\omega)}{\int_0^\infty {\rm{d}}\omega \; \omega J(\omega)}.
\end{align}
The spectral density function is a density of states weighted by the system-bath coupling energies.
The bare system's Hamiltonian is transformed into $\hat{{H}}^{\rm{eff}}_S $, which 
may depend on the parameters of the bath.
If the transformation successfully weakens the interaction between the system and bath, we may approximate the equilibrium state using a Gibbs state with respect to the effective system Hamiltonian $\hat{{H}}^{\rm{eff}}_S$.
In the EFFH framework, $\hat O$ comprises three steps as detailed in Ref.~\cite{PRXRC}:

\begin{enumerate}[(i)]
\item{Reaction coordinate (RC) mapping: This exact unitary mapping acted on Eq.~\eqref{eq:h_total_model} is performed on the bosonic reservoir by identifying a collective degree of
freedom (reaction coordinate) to be extracted from the reservoir and incorporated into the system. The resulting extended open
system  comprises the original system along with the reaction-coordinate mode. It is coupled to a so-called residual bath with a modified spectral density function \cite{Nazir18}. The goal of this transformation is to weaken the system-bath coupling strength, compared to the original model.}
\item{Polaron transformation: This unitary operation is applied on the reaction coordinate. It {\em imprints} the RC coupling into the original system and at the same time, it partially decouples the RC and the system. This step further generates new direct interaction terms between the original system and the residual bath.}
\item{Truncation: The Hamiltonian resulting from the previous step is truncated, assuming that only the ground state of the polaron-transformed reaction coordinate is populated. This approximation relies on the reaction-coordinate frequency (which derives from the spectral function of the original bath) being the largest energy scale in the problem, exceeding the thermal energy and the systems' frequencies.}
%, which is our working assumption.}
\end{enumerate}

Once the  procedure as detailed above is performed, an effective Hamiltonian emerges in the form of Eq. \eqref{eq:Heff}.
Mathematically, it resembles the original model. However, the parameters in the EFFH contain an explicit dependence on the original system-bath coupling parameters. Since the procedure is carried out in order to weaken the system-bath coupling, we conjecture
that the dynamics and steady state properties of the EFFH can be studied using weak coupling techniques such as the Redfield equation~\cite{Nitzan,Breuer}.
Altogether, the EFFH method allows studies of strong coupling thermodynamics at the cost of weak coupling approaches.

We will keep the discussion in this paper as general as possible, but to illustrate our approach we use the generalized spin-boson model, a ubiquitous model in open quantum systems. It has a wide variety of effects and applications from quantum phase transitions,
reactions dynamics, heat transport to thermometry. 
The model consists  a central spin impurity (spin splitting $2\Delta$) coupled to a bath of noninteracting harmonic oscillators. It is given by the Hamiltonian 
\bea
    \hat{H} = \Delta\hat{\sigma}_z + \sum_k \nu_k \left( \hat{c}_k^{\dagger} + \frac{t_k}{\nu_k}\hat{\sigma}_{\theta} \right) \left( \hat{c}_k + \frac{t_k}{\nu_k} \hat{\sigma}_{\theta} \right),
    \label{eq: general spin boson}
\eea
which is a special case of Eq.~\eqref{eq:h_total_model}.
In the above, the coupling operator $\hat{\sigma}_{\theta} = \cos(\theta)\hat{\sigma}_z + \sin(\theta)\hat{\sigma}_x$ 
spans between that of the usual spin-boson model at $\theta = \pi/2$ where the bath leads to relaxation dynamics and, at the other extreme, when $\theta = 0$ where the spin qubit experiences purely dephasing dynamics due to the reservoir.
In the general case, the spin dynamics reflects both population relaxation and decoherence due to its interaction with the bath.

%===================
% Figure 1
\begin{figure*}[t]
\fontsize{13}{10}\selectfont 
\centering
\includegraphics[width=1.8\columnwidth]{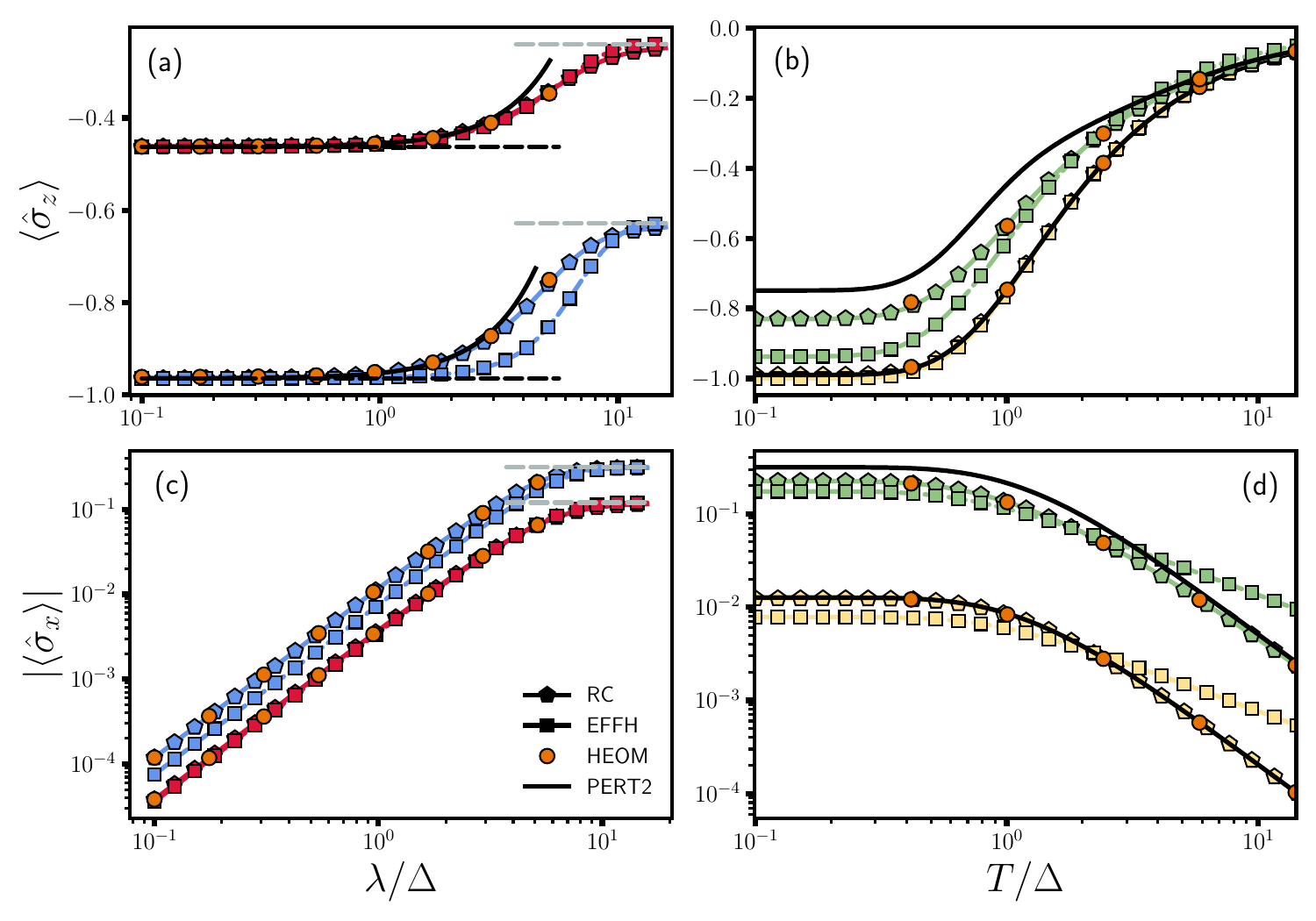}
\caption{Comparison of the equilibrium state (top) polarization and the (bottom) coherence of the spin-boson model for $\theta = \pi / 4$ as a function of the [(a),(c)] coupling strength $\lambda$ and [(b),(d)] temperature $T$. The calculations are shown for $\Omega = 8\Delta$. Pentagons denote the quantities obtained from the RC approach using $M = 50$, which is sufficient to guarantee convergence in the regime showed. Squares denote the result with the effective Hamiltonian model. Circles are computed from the hierarchical equations of motion ($N_k = 15$ and $N_c = 5$). In panels [(a),(c)], calculations are shown for two different values of the temperature. $T = 0.5\Delta$ (blue) and $2.0\Delta$ (red). In panels [(b),(d)] two values of $\lambda = \Delta$ (yellow) and $5\Delta$ (green) are chosen. The dashed lines show the values in both the weak-coupling (black-dashed) and ultrastrong coupling (grey-dashed) approximations. The solid-black lines depict results using the second order correction to the Gibbs state.
We use a Brownian spectral function with width parameter $\gamma=0.0075$. % 
}
\label{fig:HEOM}
\end{figure*}
%==============
\subsection{Benchmarking the EFFH against numerically exact results}
\label{sec:bench}

To test the accuracy of the EFFH method, we proceed to benchmark the expectation values of observables using the generalized spin-boson model [Eq.~\eqref{eq: general spin boson}] against some other approaches including numerically-exact simulations.

As for the EFFH method described before, we now recount its steps on the spin-boson model. First, the reaction coordinate mapping leads to
\begin{align}
\label{eq:HRC}
\hat{H}_{\rm RC} &= \hat{H}_S +  \Omega \left(\hat{a}^{\dagger} + \frac{\lambda}{\Omega}\hat{S} \right) \left(\hat{a} + \frac{\lambda}{\Omega}\hat{S} \right) 
\\ \nonumber
&+ \sum_k \omega_k \left(\hat{b}_k^{\dagger} + \frac{f_k}{\omega_k} (\hat{a}^{\dagger} + \hat{a}) \right) \left(\hat{b}_k + \frac{f_k}{\omega_k} (\hat{a}^{\dagger} + \hat{a}) \right),
\end{align}
where $\hat{a}$ and $\hat{a}^{\dagger}$ are canonical bosonic operators for the reaction coordinate~\cite{PRXRC}.
%The parameters $\lambda$ and $\Omega$ are calculated from the original spectral density function as 
%
%\begin{align}
%\label{eq:lambdaOmega}
%\lambda^2 = \frac{1}{\Omega} \int_0^\infty {\rm{d}}\omega \; \omega J(\omega), \quad \Omega^2 = \frac{\int_0^\infty {\rm{d}}\omega \; \omega^3 J(\omega)}{\int_0^\infty {\rm{d}}\omega \; \omega J(\omega)}.
%\end{align}
%
%The spectral density function is a density of states weighted by the system-bath coupling energies.
It is clear from Eq.~\eqref{eq:lambdaOmega} that $\lambda$ grows as the system-bath coupling is increased.
Henceforth, we thus regard $\lambda$ as the system-bath interaction energy.
The residual bath operators can be related to the operators previous to the mapping, as described in Ref.~\cite{Nazir18}

Second, we apply the polaron transformation $\hat{U}_P = {\rm{exp}}[\frac{\lambda}{\Omega}(\hat{a}^{\dagger} - \hat{a})\hat{S}]$ after the reaction coordinate mapping [step (ii) in Sec.~\ref{subsec:EFFH}]. This operation transforms Eq.~\eqref{eq:HRC} to
\begin{align}
\hat{\tilde{H}}_{\rm RC-P} &= \hat{\tilde{H}}_S + \Omega \hat{a}^{\dagger} \hat{a} \nonumber \\
&+ \sum_k \omega_k \left\{ \left[ \hat{b}_k^{\dagger} + \frac{f_k}{\omega_k} \Big( \hat{a}^{\dagger} + \hat{a} - \frac{2\lambda}{\Omega}\hat{S} \Big) \right] \right. \nonumber \\
&\times \left. \left[ \hat{b}_k + \frac{f_k}{\omega_k} \Big( \hat{a}^{\dagger} + \hat{a} - \frac{2\lambda}{\Omega}\hat{S} \Big) \right] \right\}.
\label{eq:HRCP}
\end{align}
After the truncation of the degree of freedom of the reaction-coordinate [step (iii) in Sec.~\ref{subsec:EFFH}], we find
\begin{align}
\hat{H}^{\rm{eff}} = \hat{{H}}^{\rm{eff}}_S + \sum_k \omega_k \left[ \hat{b}_k^{\dagger} - \frac{2\lambda f_k}{\Omega \omega_k} \hat{S} \right] \left[ \hat{b}_k - \frac{2\lambda f_k}{\Omega \omega_k} \hat{S} \right].
\label{eq:SBeff}
\end{align}
Here,
\bea
\hat{{H}}^{\rm{eff}}_S &= \hat{Q}_0 \hat{U}_P \hat{H}_S \hat{U}_P^{\dagger} \hat{Q}_0,
\label{eq:HSEFF}
\eea
where $Q_0 \defeq \ket{0} \bra{0}$ is the projector onto the ground-state subspace of the polaron-transformed reaction coordinate.
We note that these operations leave the form in Eq.~\eqref{eq: general spin boson} unchanged. As described in Ref.~\cite{PRXRC}, the redefined system in Eq.~\eqref{eq:SBeff} may be weakly coupled to the bath depending on the form of the original spectral function. For example, the Brownian function in Eq.~\eqref{eq:Brownian} needs to be narrow to justify weak coupling at the end of the process ($\gamma\ll1$ for the spin-boson model).
%suggesting that $\gamma$ needs to be the smallest energy parameter in the configuration.  % DS gamma is a dimensinless parameter. We had gamma\Delta <<1

A simple calculation on the spin-boson model allows us to build the effective Hamiltonian of the system. Considering  Eq.~\eqref{eq: general spin boson},
%= \cos(\theta)\hat{\sigma}_z + \sin(\theta)\hat{\sigma}_x$, 
we obtain 
\begin{align}
\hat{{H}}^{\rm{eff}}_S 
&= \frac{\Delta}{2}\left[(1+e^{-\frac{2\lambda^2}{\Omega^2}}) + (1-e^{-\frac{2\lambda^2}{\Omega^2}})\cos(2\theta)\right]\hat{\sigma}_z \nonumber\\ 
&+ \frac{\Delta}{2}\left(1-e^{-\frac{2\lambda^2}{\Omega^2}}\right)\sin(2\theta)\hat{\sigma}_x,
\label{eq:SBS}
\end{align}
which is our EFFH for the spin-boson model.
Since we work in the regime where the newly-mapped system Hamiltonian is weakly-coupled to the residual reservoir, we conjecture that the MFGS follows from the Gibbs state of the above Hamiltonian,
\begin{align}
\hat{\rho}_S = \frac{e^{-\beta \hat{{H}}^{\rm{eff}}_S}}{{\rm Tr}_S\left(e^{-\beta \hat{{H}}^{\rm{eff}}_S}\right)}.
\end{align}
Note that calculating this state is straightforward since the dimensionality of the effective system Hamiltonian is simply a two-level system.
In fact, the MFGS of the EFFH method can be obtained analytically as we showed in Ref.~\cite{PRXRC}.

%As for the assumption of weak system-bath coupling after the EFFH procedure, its validity relies on the choice of the original spectral function. For example, having $\gamma$ in Eq. \eqref{eq:Brownian} as the smallest energy parameter guarantees weak coupling after the mapping~\cite{PRXRC}.

To benchmark the ability of the EFFH method to describe the MFGS, we compute expectation values $\langle \hat{O} \rangle = \Tr[\hat{\rho}_S \hat{O}]$. For the spin-boson model, both $\hat{O} = \hat{\sigma}_x$ and $\hat{O} = \hat{\sigma}_z$ completely determine the state. 
As for the spectral function of the bath, we work with the Brownian spectral function  [Eq.~\eqref{eq:Brownian}] since it can be effectively simulated with alternative numerical techniques that will serve as benchmarks.

In Fig.~\ref{fig:HEOM} we display both $\langle \hat{\sigma}_z \rangle$ and $\langle \hat{\sigma}_x \rangle$ as a function of the coupling $\lambda$ at fixed temperature $T$ [Fig.~\ref{fig:HEOM}(a),(c)] and as a function of $T$ at fixed $\lambda$ [Fig.~\ref{fig:HEOM}(b),(d)]. 
We compare  the equilibrium state obtained from the effective Hamiltonian treatment with results from the HEOM method, a numerically-exact approach. Appendix~\ref{app:1} provides technical details on the HEOM simulations.
We further display simulations from the  reaction-coordinate (RC) technique~\cite{NickRC, NickRC2}. This amounts to constructing  the Gibbs state from the original system Hamiltonian $\hat{H}_S$ together with the degrees of freedom of the reaction coordinate [first line in Eq.~\eqref{eq:HRC}], followed by a partial trace over the degrees of freedom of the reaction coordinate.
We remark that, for our chosen parameters, converging the HEOM equations becomes increasingly difficult as $\lambda$ increases, particularly in the low temperature regime. Nevertheless, it can be observed that for the results shown in Fig.~\ref{fig:HEOM} the reaction coordinate perfectly agrees with HEOM simulations. On the other hand, the EFFH approximation coincides with these results depending on both $\lambda$ and $T$. 
We also show in Fig.~\ref{fig:HEOM} results based on the 2$^{\rm nd}$ order perturbation scheme for the MFGS as presented in Ref.~\cite{JanetPRL}; details of this calculation are provided in Appendix~\ref{app:2}. As expected, this approximation is only valid for moderate coupling strength $\lambda$.
Limiting cases are also marked in Fig.~\ref{fig:HEOM} with dashed lines for the ultraweak coupling limit and ultrastrong coupling limits. 
Focusing our attention on the coupling-strength dependence of both polarization and coherences, we make the following observations based on Fig.~\ref{fig:HEOM}: 

(i) RC simulations perfectly agree with numerically exact results. We further verified this correspondence for a smaller characteristic frequency of the bath, $\Omega=1$ (not shown).  Below, this excellent agreement allows us to test the EFFH method against RC simulations, rather than perform the computationally-costly HEOM calculations.

(ii) The EFFH treatment provides accurate results at both the ultraweak and ultrastrong coupling limits. Furthermore, it is correct quantitatively at all coupling regimes as long as the
temperature is comparable to the spin splitting $\Delta$. 
The EFFH method struggles in the intermediate coupling regime either when the temperature is very low $T\ll \Delta$ or high $T\gtrsim\Omega$.
In the former case,
one requires the residual coupling to the bath to be sufficiently weak for the effective Hamiltonian treatment to provide accurate results for the MFGS. We can argue that at low $T$,  correlations between the system and the reservoir are more prominent. This requires a more detailed description than our effective Hamiltonian for the physical effects to be captured. As for high temperatures, the extreme truncation of the RC limits the applicability of the EFFH method to the regime of $\Omega\gg T$.
 Looking at $\langle \hat{\sigma}_x \rangle$ in Fig.~\ref{fig:HEOM}[(c),(d)], we particularly note that the coherence as calculated from the EFFH method are underestimated at low temperature and overestimated at high $T$. %This may be explained from the nature of our truncation scheme, as higher values of $T$ imply that excited states of the reaction coordinate acquire larger occupation values that are not captured via our ground-state truncation. % DD2 this does not explain why we overestimate coherences

(iii) The second-order correction to the Gibbs state only provides an accurate representation of the equilibrium state at weak-to-moderate coupling $\lambda$, while our effective Hamiltonian model is able to capture the full range of coupling.
It is intriguing that the effective Hamiltonian model describes the ultrastrong coupling limit so well; this result was discussed in Ref. \cite{PRXRC}.

Summarizing our benchmark efforts, we conclude that the EFFH method is a highly-reliable tool in situations when the residual coupling of the system to the bath is weak. It provides quantitatively-correct results to the MFGS at any coupling for temperatures in the range $\Delta \approx T \ll \Omega$. 
%and when $\lambda\ll \Omega$. % DD2 commented. 
Significantly, with its minimal computational effort, the EFFH approach produces results on par with numerically-exact simulations.

%======================================
%\section{The variational Reaction Coordinate Polaron Transformation method}
\section{The Variational EFFH (var-EFFH) method}
\label{sec:vEFFH}

\subsection{Presentation of Method}
\label{sec:veff}

Due to the truncation of the polaron-transformed reaction coordinate harmonic ladder to the {\it ground} state, the EFFH method is limited to handle situations in which $\Omega\gg T,\Delta$. In physical terms, the collective bath's degree of freedom to which the system is strongly coupled needs to be of high frequency relative to other energy scales in the system.

In this Section, we introduce a variational EFFH (var-EFFH) method, extending the original EFFH framework of Ref.~\cite{PRXRC} to handle the regime in which $\Omega \approx T,\Delta$.
The outcome of the var-EFFH method is an effective Hamiltonian that depends on both the system-bath energy parameters and the bath's temperature. Simulations presented in Sec.~\ref{sec:simul} demonstrate that this tool indeed can capture the behavior of polarization of the spin-boson model when $\Omega$ is comparable to $\Delta$.

% where a plain polaron transformation was implemented. 

%We  extract a collective mode of frequency $\Omega$ from the reservoir. It couples to the system at a strength $\lambda$. This standard step results in the RC Hamiltonian
%
%\bea
%\label{eq:h_rc_tot}
%\hat{H}_{\rm{RC}} &=& \hat{H}_S + \Omega \Big( \hat{a}^{\dagger} + \frac{\lambda}{\Omega}\hat{S} \Big) \Big( \hat{a} + \frac{\lambda}{\Omega}\hat{S} \Big) \\
%&+& \sum_k \omega_k \left[ \hat{b}^{\dagger} + \frac{f_k}{\omega_k} (\hat{a}^{\dagger} + \hat{a}) \right] \left[ \hat{b} + \frac{f_k}{\omega_k} (\hat{a}^{\dagger} + \hat{a}) \right].
%\nonumber
%\eea
%
We recall the three steps in deriving the effective Hamiltonian, as summarized in Sec.~\ref{subsec:EFFH}.
The first step in the protocol involves performing a reaction coordinate transformation. In the second step,
a polaron transformation is applied on the RC.
In the variational EFFF approach, rather than applying a ``normal" (or complete) polaron transform, we
apply it in a variational manner defining
\bea
\hat{\tilde{H}} = \hat{U}_P \hat{H}_{\rm{RC}} \hat{U}^{\dagger}_P, 
\eea
with $\hat{U}_P = {\rm{exp}}[\eta(\hat{a}^{\dagger} - \hat{a})\hat{S}]$. Here, $\eta$ is a real-valued variational parameter to be optimized later through the minimization of the free energy, based on the effective Gibbs state being thermal \cite{Cao2}, leading to $\eta \le \lambda/\Omega$.
The $\hat{U}_P$ transformation shifts the RC operator according to $\hat a \to \hat a - \eta \hat S$.
The total Hamiltonian is given by
\bea
\nonumber
\hat{\tilde{H}}_{\rm RC-P} &= &\hat{\tilde{H}}_S + \Omega \left [\hat{a}^{\dagger} + \left(\frac{\lambda}{\Omega} - \eta\right)\hat{S} \right] 
\left[ \hat{a} + \left(\frac{\lambda}{\Omega} - \eta\right)\hat{S} \right]  \\
&+& \sum_k \omega_k \left\{ \left[ \hat{b}^{\dagger} + \frac{f_k}{\omega_k} \Big( \hat{a}^{\dagger} + \hat{a} - 2\eta\hat{S} \Big) \right] \right. \nonumber \\
&\times& \left. \left[ \hat{b} + \frac{f_k}{\omega_k} \Big( \hat{a}^{\dagger} + \hat{a} - 2\eta\hat{S} \Big) \right] \right\}.
\eea
This Hamiltonian replaces Eq.~\eqref{eq:HRCP}.
As we show below, $\eta$ is in  general a function of  the parameters of the system and its environment, including the temperature. 

The third step in the procedure is to project this Hamiltonian onto a subspace in which the RC occupies only its ground state, $\ket{0}$. This is achieved using the operator $\hat{Q}_0 = \ket{0}\bra{0}$. The result is the variational-effective Hamiltonian, and we highlight its dependence on $\eta$,
\bea
\hat{H}^{\rm{Veff}}(\eta) &=& \hat{Q}_0 \hat{\tilde{H}}_S \hat{Q}_0 + \Omega \left(\frac{\lambda}{\Omega} - \eta \right)^2 \hat{S}^2
 \nonumber\\
&+& \sum_k \omega_k \left[ \hat{b}^{\dagger} - \frac{2\eta f_k}{ \omega_k} \hat{S} \right] \left[ \hat{b} - \frac{2\eta f_k}{ \omega_k} \hat{S} \right].
\label{eq:HVEFF}
\eea
%
%Recall that $\lambda$ corresponds to the system-bath coupling parameter. 
%The impact of strong coupling is thus now embedded in the parameter $C$ dressing the system portion of the Hamiltonian. For problems of thermalization, we thermalize now with respect to the 
We now identify
the system's contribution to the var-EFFH, given by 
\bea
    \hat{H}^{{\rm Veff}}_S(\eta) = \hat{Q}_0 \hat{\tilde{H}}_S \hat{Q}_0 + \Omega \left(\frac{\lambda}{\Omega} - \eta \right)^2 \hat{S}^2.
\eea
% XXX
As expected, when performing the full polaron transform using $\eta = \frac{\lambda}{\Omega}$, the second term disappears and we recover Eq.~\eqref{eq:HSEFF}.
The bath Hamiltonian is simply $\hat H_B=\sum_k\omega_k \hat b^{\dagger}_k \hat b_k$, and the system-bath coupling is given by
$\hat H_I=-2\eta\hat S\sum_k\frac{f_k}{\omega_k} \left(\hat b^{\dagger} + \hat b\right)$.

To find the parameter $\eta$, we minimize the Gibbs-Bogoliubov-Feynman upper bound on the free energy~\cite{Silbey,Cao2} given by the following expression
\begin{equation}
\label{eq:GBF upperbound}
    A_B = -\frac{1}{\beta}\ln \Tr\left[e^{-\beta[\hat{H}^\text{Veff}_S(\eta)+\hat{H}^\text{Veff}_B]}\right]+\langle\hat{H}^\text{Veff}_I\rangle_{\hat{H}^\text{Veff}_S(\eta)+\hat{H}^\text{Veff}_B}.
\end{equation}
Here, the average is done over a canonical state with respect to the written Hamiltonian.
Since $\langle\hat{H}^\text{Veff}_I\rangle_{\hat{H}^\text{Veff}_S(\eta)+\hat{H}^\text{Veff}_B}=0$ by construction, and the bath Hamiltonian is independent of the parameter $\eta$, we simply solve for $\eta$ using
\begin{equation}
    \frac{\partial}{\partial \eta}\left[-\frac{1}{\beta}\ln \Tr_S\left(e^{-\beta\hat{H}^\text{Veff}_S(\eta)}\right)\right]=0.
\end{equation}
%we must optimize with respect to this parameter under the assumption that the equilibrium state is as Gibbs ensemble with respect to the variational effective system Hamiltonian at inverse temperature $\beta$. This can be done by computing the lower bound on the free energy $A_{\beta}$. Since the reservoir is composed of non-interacting harmonic oscillators, we obtain the following equation to optimize
%\bea
%\frac{d}{dC} \ln\left( \Tr \left[ e^{-\beta\hat{H}_S^{Veff}(C)} \right]\right) = 0.
%\eea
This equation readily simplifies to
\bea
\Tr_S \left[  e^{-\beta\hat{H}_S^{{\rm Veff}}(\eta)}\frac{d}{d\eta} \hat{H}_S^{{\rm Veff}}(\eta)\right] = 0.
\label{eq:Veffcond}
\eea
To compute the left hand side, one can build a closed-form expression for $\hat{H}_S^{{\rm Veff}}(\eta)$ for the particular model and set up a transcendental equation for $\eta$, which is then solved numerically. This approach is feasible if $\hat{S}^2 \propto \hat{I}$ since computing the exponential term analytically can be a difficult task. An alternative approach is to solve the problem fully numerically by calculating the derivative  of $\hat{H}_S^{{\rm Veff}}(\eta)$ with respect to $\eta$ then implementing the routine numerically. 
Once the optimal value of $\eta$ (denoted as $\eta_{v}$) has been obtained, the thermal state of the system is approximated as 
\bea
\hat{\rho}^{\rm Veff}_S =
 \frac{e^{-\beta\hat{H}_S^{{\rm Veff}}(\eta_v)} } { \Tr_S \left[e^{-\beta\hat{H}_S^{{\rm Veff}}(\eta_v)}\right]  },
\eea
where we emphasize again, $\eta_v = \eta(\lambda,\Omega,T,... )$.
% How do we know we minimized?... 
%====================================
% Figure 2
\begin{figure*}
    \centering
\includegraphics[width=2\columnwidth]
{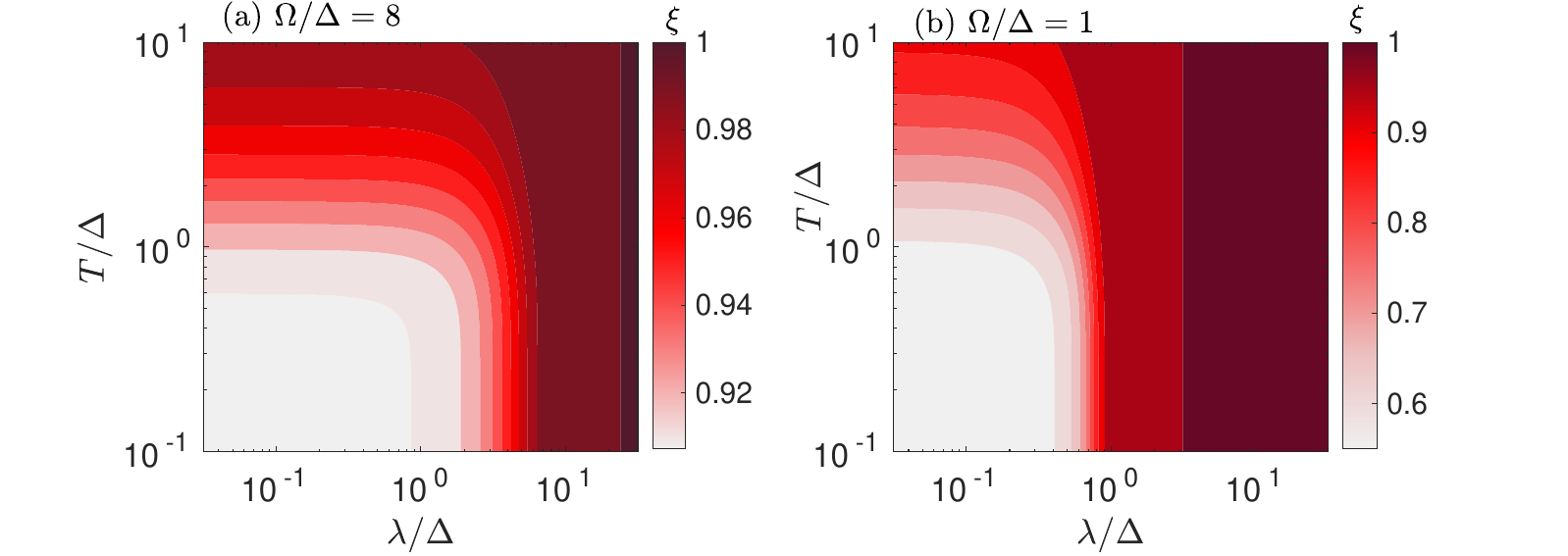}%{fig2N.eps}
\caption{The variational parameter $\xi$
for the generalized spin-boson model as a function of 
coupling strength $\lambda$ and temperature $T$. 
This parameter, defined from the relation,
$\eta_{v}=\xi\frac{\lambda}{\Omega}$,
 satisfies $0\leq\xi\leq1$.
 When $\xi\to1$, the variational treatment is redundant and the original EFFH method can be used.
(a) $\Omega/\Delta = 8$, 
(b) $\Omega/ \Delta=1$.
Other parameters are $\Delta = 1$ and $\theta = \pi/4$.
}
\label{fig:fig1}
\end{figure*}
%====================================

\subsection{Application to the generalized spin-boson model}

Following the var-EFFH procedure described in Sec.~\ref{sec:veff}, we write down the corresponding var-EFFH system Hamiltonian for the spin-boson model,
\bea
\nonumber
\hat{H}^{\rm{Veff}}_{S}(\eta)&=& \frac{\Delta}{2}\left[(1+e^{-2\eta^2}) + (1-e^{-2\eta^2})\cos(2\theta)\right]\hat{\sigma}_z
\\ \nonumber
&+& \frac{\Delta}{2}\left(1-e^{-2\eta^2}\right)\sin(2\theta)\hat{\sigma}_x + \Omega \left( \frac{\lambda}{\Omega} - \eta \right)^2.
\\
\label{eq:HSBVeff}
\eea
To find the optimal value of $\eta$ we first compute
\bea
\frac{d}{d\eta} \hat{H}^{\rm{{\rm Veff}}}_{S}(\eta) &=& -
2\Omega\left(\frac{\lambda}{\Omega} - \eta\right) 
\\ \nonumber
&+& 2\Delta \eta e^{-2\eta^2} \left[ (\cos(2\theta)-1 )\hat{\sigma}_z + \sin(2\theta)\hat{\sigma}_x\right].
\eea
Then, using properties of Pauli matrices we analytically solve this problem from Eq.~\eqref{eq:Veffcond}, revealing a transcendental equation of the form,
\bea
    \frac{\lambda}{\Omega} = \eta_v\left[ 1 + \frac{2\Delta e^{-4\eta_v^2} (1-\cos 2\theta)}{\Omega |\Vec{v}|} \tanh\left( \frac{\beta\Delta}{2}|\Vec{v}|\right)\right].
    \nonumber\\
    \label{eq:C1}
\eea
In the above expression, $|\Vec{v}|$ depends on $\theta$ and $\eta_v$ as
\bea
|\Vec{v}| = \sqrt{ 2(1+e^{-4\eta_v^2}) + 2(1-e^{-4\eta_v^2})\cos 2\theta  }.
\eea
Setting $\eta_v \equiv \xi\frac{\lambda}{\Omega}$, which implies that $0 \le \xi \le 1$,  Eq.~\eqref{eq:C1} transforms into 
\bea
    1 = \xi\left[ 1 + \frac{2\Delta e^{-4\xi^2\frac{\lambda^2}{\Omega^2}} (1-\cos 2\theta)}{\Omega |\Vec{v}|} \tanh\left( \frac{\beta\Delta}{2}|\Vec{v}|\right)\right].
    \nonumber\\
    \label{eq:C2}
\eea
 It is clear that the effective parameter $\eta_v$ depends not only on the system-bath characteristic interaction energy scale $\lambda$ and frequency $\Omega$, but also on the bath temperature,  extending the results of Ref.~\cite{PRXRC}.
We now study particular cases and show how the variational parameter $\eta_v$  behaves in interesting limits. 

(i) {\it Ultrastrong coupling limit, $\lambda \xrightarrow{} \infty$}. In this case,  we get from Eq.~\eqref{eq:C2} that
\bea
    \xi = 1,
\eea
which implies that the variational approach is redundant in the ultrastrong coupling limit.
This is to be expected since we know that the EFFH method of Ref.~\cite{PRXRC} is exact for the spin-boson model
in the ultrastrong coupling limit, as observed in Sec.~\ref{sec:bench}. 
% 
%More deeply,  the fact that the variational approach does not add corrections to the MFGS in the ultrastrong coupling limit corresponds to the temperature not being an important parameter in this limit, given that properties of the MFGS are controlled predominantly by the coupling energy.

(ii) {\it Asymptotically weak coupling $\lambda \xrightarrow{} 0$}. In this regime,
\bea
    \xi &\approx& \frac{1}{1 + \frac{\Delta}{\Omega} [1 - \cos2\theta] \tanh(\Delta \beta)}
    \nonumber\\
    &\approx& 1 - \frac{\Delta}{\Omega} [1 - \cos2\theta] \tanh(\Delta \beta),
\eea
derived once requiring that $\Omega$ is the largest energy scale in the problem. Note that 
the correction to unity is small as a function of temperature since  $\tanh\beta \Delta$ extends from $\beta \Delta$ at high temperature to one at very low temperatures.

(iii) {\it Low temperature, $\beta \xrightarrow{} \infty$}. In this case we find from Eq.~\eqref{eq:C2} that
\bea
    1 = \xi\left[1 + \frac{2\Delta e^{-4\xi^2\frac{\lambda^2}{\Omega^2}} (1-\cos 2\theta )}{\Omega |\Vec{v}|}\right].
\eea
In this regime, the temperature only minimally impacts the variational parameter. 

(iv) {\it High temperature $\beta \xrightarrow{} \infty$}. In this limit we find that 
\bea
    1 = \xi\left[1 + \beta \frac{\Delta^2 e^{-4\xi^2\frac{\lambda^2}{\Omega^2}} (1-\cos2\theta)}{\Omega } + \mathcal{O}(\beta^3)\right].
\eea

In Fig.~\ref{fig:fig1} we present simulations of $\xi$ as a function of system bath-coupling strength and temperature for $\theta= \pi/4$. 
We find that for $\Omega\gg T,\Delta$, the variational parameter $\xi$ does not reduce below 0.88
throughout the full range of coupling energies and temperatures.
Thus, the expected correction due to the variational treatment is not large in this regime and the regular-nonvariational EFFH method is sufficient. 
On the other hand, for small values of the bath's spectral frequency, $\Omega/\Delta=1$,
corrections can be as substantial as 0.5.

We note that the variational approach corrects the EFFH method in the {\it low temperature} and {\it weak-to-intermediate} coupling regime. 
In the strong coupling limit and at high temperatures, we find that  $\xi\approx 1$, thus one need not be concerned with these corrections.

The modification to the spin-boson Hamiltonian appears with  the renormalization factor that we now define as
\bea
\kappa_{\rm E} \equiv e^{-2\xi^2 \frac{\lambda^2}{\Omega^2}}.
\label{eq:kappaEff}
\eea
Using $\xi=1$, this expression reduces to the non-variational treatment of Ref. \cite{PRXRC}. 

%===========================
\section{Comparative analysis: The variational polaron MFGS} 
\label{sec:polaron}

The polaron transform is a well-established and popular treatment to study strong system-bath coupling effects in spin-boson type models. In particular, it has been extensively employed to study heat transfer in nonequilibrium settings where a few level system(s) are coupled to two or more reservoirs~\cite{diodeSC0, diodeSC1,  diodeSC2,diodeSC1b,Pollock_2013,Hsieh_2019,Zhang_2015,Zhang_2015_2,Liu_2018,Wang_2015,Liu_2017,Koch_2011,Dominguez_2011,Lu_2019,Wachtler_2020,Koch_2010,Chen_2019,Wang_2017}. Furthermore, the polaron transform has also been utilized to analytically and numerically gain insights on experimental works on superconducting circuits~\cite{Dai_2022,Agarwal_2013,Shi_2018}. Non-Markovian dynamics in cavity quantum electrodynamics have been also examined using polaron transformed QME~\cite{Zhangg_2022,bn_2021}. Critical phenomena in dissipative models~\cite{Kopylov_2019} and the robustness of topological order in two-dimensional spin lattice coupled to bosonic reservoir~\cite{Pedrocchi_2013} are other recent examples where the transform was  found useful. In the context of quantum thermodynamics~\cite{JanetRev,Cao1,Cao2,Cao3}, although the polaron transform was used to build the MFGS in the intermediate system-bath coupling regime, the discussion has been mostly limited to the spin-boson model.

In what follows, we first introduce the procedure of obtaining an approximate MFGS for a {\it general} open system using the variational polaron transformation.
We then exercise this method on the generalized spin-boson model described in 
Eq.~\eqref{eq: general spin boson}. 
We discuss the relationship of the MFGS as obtained from the variational polaron (var-POL) to the MFGS of the  var-EFFH method.
 Finally, we present simulations for the spin polarization and coherences for baths with spectral function peaked either at high or low frequencies.

%===============================================
\subsection{Polaronic MFGS: Generic open quantum system}

The variational polaron MFGS  was constructed for the spin-boson model and benchmarked against other methods \cite{Cao1,Cao2}. Here,
we present the generalization of the procedure to obtain the MFGS for a generic open quantum system as in Eq.~\eqref{eq:generic}. The polaron transform is generated here by the following unitary,
\bea
 \hat{W}=\exp(-\ii\hat{S}\hat{B}/2),
 \eea
where $\hat{B}=2\ii\sum_k\frac{f_k}{\nu_k}(\hat{c}^\dagger_k-\hat{c}_k)$ and $\hat S$ is the system operator coupled to the bath, introduced in Eq.~\eqref{eq:generic}.
 The transformation is referred to as ``full-polaron" if the variational parameters $\{f_k\}$ are simply set to $\{t_k\}$, the original system-reservoir couplings. If, instead, the optimal values for $\{f_k\}$ are obtained by minimizing the Gibbs-Bogoliubov-Feynman upper bound on the free energy, the transform is called ``variational" \cite{Silbey,Silbey_2}. 
 Performing the transform  on the Hamiltonian Eq.~\eqref{eq:generic} we obtain
\begin{equation}
\begin{aligned}
\hat{H}^\text{pol} = &\hat{W} \hat{H}_S \hat{W}^\dagger+\hat{S}^2\sum_k\frac{(t_k-f_k)^2}{\nu_k}+\hat{H}_B\\
+&\hat{S}\sum_k(t_k-f_k)\left(\hat{c}^\dagger_k+\hat{c}_k\right),
    \end{aligned}
\end{equation}
with $\hat{H}_B=\sum_k \nu_k c^\dagger_kc_k$. 

The next stage in building a Hamiltonian of mean force is to add and subtract the bath-averaged expectation value of the transformed system Hamiltonian. This ensures that the thermal average of the interaction operator diminishes. We thus write down the above total Hamiltonian as
%%
%\begin{equation}
%\begin{aligned}
%\hat{H}^\text{pol} = &\hat{W} \hat{H}_S \hat{W}^\dagger+\langle\hat{W} \hat{H}_S \hat{W}^\dagger\rangle_{\hat{H}_B}-\langle\hat{W} \hat{H}_S \hat{W}^\dagger\rangle_{\hat{H}_B}\\
%+&\hat{S}^2\sum_k\frac{(t_k-f_k)^2}{\nu_k}+\hat{H}_B\\
%+&\hat{S}\sum_k(t_k-f_k)\left(\hat{c}^\dagger_k+\hat{c}_k\right),
 %   \end{aligned}
%\end{equation} 
% 
%We now identify the system, bath, and the system bath coupling terms in 
\bea 
\hat{H}^\text{pol} = \hat{H}^\text{pol}_S+\hat{H}^\text{pol}_B+\hat{H}^\text{pol}_I,
\eea 
identifying the system, bath, and the interaction terms, respectively, by
%as follows,
%with or the transformed system, bath, and the interaction part of the Hamiltonian, we obtain:
%
\bea
    \hat{H}^\text{pol}_S &=& \langle \hat{W}  \hat{H}_S \hat{W}^\dagger\rangle_{\hat{H}_B}+\hat{S}^2\sum_k\frac{(t_k-f_k)^2}{\nu_k}, 
        \nonumber\\
    \hat{H}^\text{pol}_B &=&\hat{H}_B, 
        \nonumber\\
    \hat{H}^\text{pol}_I &=& \hat{W}  \hat{H}_S \hat{W}^\dagger-\langle \hat{W}  \hat{H}_S \hat{W}^\dagger\rangle_{\hat{H}_B}
        \nonumber\\
    &+&\hat{S}\sum_k(t_k-f_k)\left(\hat{c}^\dagger_k+\hat{c}_k\right).
        \label{eq:HP}
    \eea
The system's Hamiltonian $\hat{H}^{\rm pol}_S$ includes two terms. First, the original system Hamiltonian is rotated to the polaron frame. Second, a new term emerges, related to the original system-bath coupling operator. This second term in particular could realize bath-induced interactions between individual impurities (e.g., spins) immersed in a bath. 
We now make the assumption that the polaron-transformed Hamiltonian is weakly coupled to the bath, thus the MFGS can be approximated from the system Hamiltonian as
\begin{equation}
    \hat{\rho}^{\rm pol}_S = \frac{e^{-\beta \hat{H}^{\rm pol}_S}}{\Tr\left(e^{-\beta \hat{H}^{\rm pol}_S}\right)}.
    \label{eq:MFGSpol}
\end{equation}
We note that it is often useful to diagonalize the system operator $\hat{S}$ that couples to the bath modes prior to the polaron transform. This is indeed the case in the general spin-boson model where $\hat{S}=\cos(\theta)\hat{\sigma}_z+\sin(\theta)\hat{\sigma}_x$. Rotating $\hat{S}$ to $\hat{\sigma}_z$ via the rotation matrix $\hat{R}=\exp(-\frac{\ii}{2}\theta\hat{\sigma}_y)$ allows us to compute the nested commutator involved in $\langle \hat{W}\hat{H}_S\hat{W}^\dagger\rangle_{\hat{H}_B}$ with more ease.
Once the transformed Hamiltonian is written down, the optimal values for $\{f_k\}$ are obtained by minimizing the free energy given by 
an expression analogous to Eq.~\eqref{eq:GBF upperbound},
\begin{equation}
    A_B = -\frac{1}{\beta}\ln \Tr\left[e^{-\beta(\hat{H}^{\rm pol}_S+\hat{H}^{\rm pol}_B)}\right]+\langle\hat{H}^{\rm pol}_I\rangle_{\hat{H}^{\rm pol}_S+\hat{H}^{\rm pol}_B}.
\end{equation}
Since $\langle\hat{H}^{\rm pol}_I\rangle_{\hat{H}^{\rm pol}_S+\hat{H}^{\rm pol}_B}=0$ by construction, and the bath Hamiltonian is independent of the parameters $\{f_k\}$, the minimization of the free energy requires solving
\begin{equation}
    \frac{\partial}{\partial f_k}\left[-\frac{1}{\beta}\ln \Tr_S\left(e^{-\beta\hat{H}^{\rm pol}_S}\right)\right]=0.
\end{equation}
Once the set of parameters $f_k$ is obtained, it is used in Eq.~\eqref{eq:HP} to build the system's Hamiltonian, and then proceed to compute the MFGS.
%===================================================
\subsection{Application to the spin-boson model} 
We apply the var-POL approach on  the general spin-boson model and construct the approximate MFGS. Applying Eq.~\eqref{eq:HP} on the model
\eqref{eq: general spin boson},  we write down
%Starting from Eq. \eqref{eq: general spin boson} and performing the variational polaron transform we get
%
$\hat{H}^{\rm pol}=E_0\hat{I}+\hat{H}^{\rm pol}_S+\hat{H}^{\rm pol}_B+\hat{H}^{\rm pol}_I$ with
\bea
%\label{eq: var pol H full}
%\begin{aligned}
    \hat{H}^{\rm pol}_S &= &\Delta\left[\cos^2(\theta)+\sin^2(\theta)\kappa_P\right]\hat{\sigma}_z
    \nonumber\\
    &+&\Delta\cos(\theta)\sin(\theta)(1-\kappa_P)\hat{\sigma}_x,
    \label{eq:HSP}
  \\
    \hat{H}^{\rm pol}_B &=& \sum_k \nu_k \hat{c}^\dagger_k \hat{c}_k,
     \label{eq:HSB}
  \\
    \hat{H}^{\rm pol}_I&=& \hat V_x \hat{\sigma}_x+\hat V_y\hat{\sigma}_y+\hat V_z\hat{\sigma}_z.
        \label{eq:HSBP}
  %  \end{aligned}
\eea
Here, $\hat{H}^{\rm pol}_S$ is obtained by first diagonalizing the system operator  $\hat{S}$
%$=\cos(\theta)\hat{\sigma}_z+\sin(\theta)\hat{\sigma}_x$ 
via the rotation matrix $\hat{R}=\exp(-\frac{\ii}{2}\theta\hat{\sigma}_y)$. We then perform the var-POL transform with the new interaction term, being the $\hat{\sigma}_z$ operator. Once we construct the new system Hamiltonian, we rotate it back to the original basis via $\hat{R}^\dagger$. $E_0 = \sum_k\frac{(f_k-t_k)^2}{\nu_k}$ is a trivial energy shift.
The system's Hamiltonian depends on $0\leq\kappa_P\leq 1$, which is given by
\bea
  \kappa_P = \exp\left[-2\sum_k\frac{f^2_k}{\nu^2_k}\coth\left(\frac{\beta\nu_k}{2}\right)\right].
  \label{eq:kappa}
  \eea
To be concrete, at specific angles of interest the system's Hamiltonian is given by
\begin{align}
    \hat{H}^{\rm pol}_S(\theta=0) &=\Delta \hat{\sigma}_z     \label{eq:HSBP0}, \\
    \hat{H}^{\rm pol}_S(\theta=\pi/2) &=\Delta \kappa_P \hat{\sigma}_z,\\
    \hat{H}^{\rm pol}_S(\theta=\pi/4) &=
    \frac{\Delta}{2}\left[\left(1+\kappa_P\right)\hat{\sigma}_z+\left(1-\kappa_P\right)\hat{\sigma}_x\right].
    \label{eq:HSBP4}
\end{align}
The angle $\theta=0$ corresponds to the exactly-solvable pure decoherence model. 
For $\theta=\pi/2$, we retrieve the standard spin-boson model with renormalized spin splitting $ \Delta  \kappa_P$. 
In between, for $\theta=\pi/4$, the polaron-transformed system's Hamiltonian includes both spin splitting and tunneling between levels. 

It can be readily shown that Eq.~\eqref{eq:HSP}
has the exact same form as the EFFH for the same system, as shown in Eq.~\eqref{eq:HSBVeff}.
The two approaches thus build a similar approximation for the MFGS with $\kappa_P$ of the var-POL mirroring the $\kappa_E$ factor  appearing in the var-EFFH Hamiltonian, Eq.~\eqref{eq:kappaEff}.
However, calculating $\kappa_P$ is significantly more complicated numerically than evaluating $\kappa_E$, since the optimization process in the former involves many (in principle infinite) modes, leading at times to convergence problems at strong coupling, as we show below in simulations.

As for the system-bath coupling operators, the bath's operators coupled to the system are notably more complicated in the polaron approach than those showing up in Eq.~\eqref{eq:HSBVeff}. In the polaron-transformed spin-boson model we obtain 
\begin{widetext}
\bea
%    E_0 =& \sum_k\frac{(f_k-t_k)^2}{\nu_k}\\
%    \kappa =& \exp\left(-2\sum_k\frac{f^2_k}{\nu^2_k}\coth(\frac{\beta\nu_k}{2})\right)\\
    \hat V_x &=& \sin(\theta)\sum_k(t_k-f_k)(\hat{c}^\dagger_k+\hat{c}_k)
%    \nonumber\\
    -\Delta\sin(\theta)\cos(\theta)\left[\cos(\hat{B})-\kappa_P\right],
    \nonumber\\
    \hat V_y &= &-\Delta \sin(\theta)\sin(\hat{B}), \nonumber\\
   \hat V_z &= & \cos(\theta)\sum_k(t_k-f_k)(\hat{c}^\dagger_k+\hat{c}_k)
   %\nonumber\\
   +\Delta \sin^2(\theta)\left[\cos(\hat{B})-\kappa_P\right].
    %\nonumber\\
    \label{eq:HSBPl}
\eea
These terms however do not impact the MFGS as constructed from Eq.~\eqref{eq:MFGSpol}.
We recall that the parameters $\{f_k\}$ are obtained through a variational approach.
As detailed in Appendix.~\ref{app:3}, we rewrite the relationship between the physical coupling energies $\{t_k\}$ and the sought-after parameters $\{f_k\}$ through a self-consistent set of equations  
\bea
f_k=F(\nu_k)t_k,
\label{eq:tf}
\eea
where, for the selected models, we get
%
%\begin{widetext}
\bea
\label{eq: F pi half and fourth}
   && F(\nu_k, \theta=\pi/2) = \left[1+\tanh\left(\beta\Delta \kappa_P\right)\coth\left(\frac{\beta\nu_k}{2}\right)\frac{2\Delta \kappa_P}{\nu_k}\right]^{-1},
    \nonumber\\
&&     F(\nu_k,\theta=\pi/4) = 
   %  \nonumber\\  
     \Bigg[1+\tanh\left(\beta\Delta\sqrt{\frac{1+\kappa_P^2}{2}}\right)
    \coth\left(\frac{\beta \nu_k}{2}\right)
    \sqrt{\frac{2}{1+\kappa_P^2}}\frac{\Delta\kappa_P^2}{\nu_k}\Bigg]^{-1}.
     %\nonumber\\
     \label{eq:F}
\eea
\end{widetext}
The parameter $\kappa_P$ is defined in Eq.~\eqref{eq:kappa}.
In practice, the numerical procedure to solve for the set of variational parameters $\{f_k\}$ involves 
first constructing a finite-truncated set of system-bath coupling energies $\{t_k\}$ based on the model spectral function and initializing $\kappa_P$ as a guess. Then, we calculate $\{f_k\}$ using Eqs.~\eqref{eq:tf}-\eqref{eq:F}, regenerate the dressing parameter $\kappa_P$ in Eq.~\eqref{eq:kappa}, and iterate this process until convergence.

At the technical level, the main differences between the two procedures, the var-EFFH mapping and the var-POL, are that in the former, we first extract a collective degree of freedom from the bath (reaction coordinate), then rotate it individually to the polaron frame.
In contrast, in the var-POL framework a polaron transform is operated
directly on the original Hamiltonian, affecting all modes in the bath.
As for the resulting Hamiltonian,  the  distinction lies in the interaction terms:  although the var-POL produces a system Hamiltonian analogous to that from the EFFH method, interaction terms are  modified. This difference between the methods does not show up in the MFGS, but it will be displayed in the study of dynamics, e.g., via the usual weak-coupling master equation technique. Concretely, in case of the general spin-boson model, the var-POL transform will induce all $\hat{\sigma}_x$, $\hat{\sigma}_y$, and $\hat{\sigma}_z$ couplings with the bath displacement operators as described by Eq.~\eqref{eq:HSBP}
and Eq.~\eqref{eq:HSBPl}. This is in contrast to the EFFH which  retains its original coupling structure, see Eq.~\eqref{eq:HVEFF}. 

%===========
% Figure 3
\begin{figure}
    \centering
\includegraphics[width=1\columnwidth]{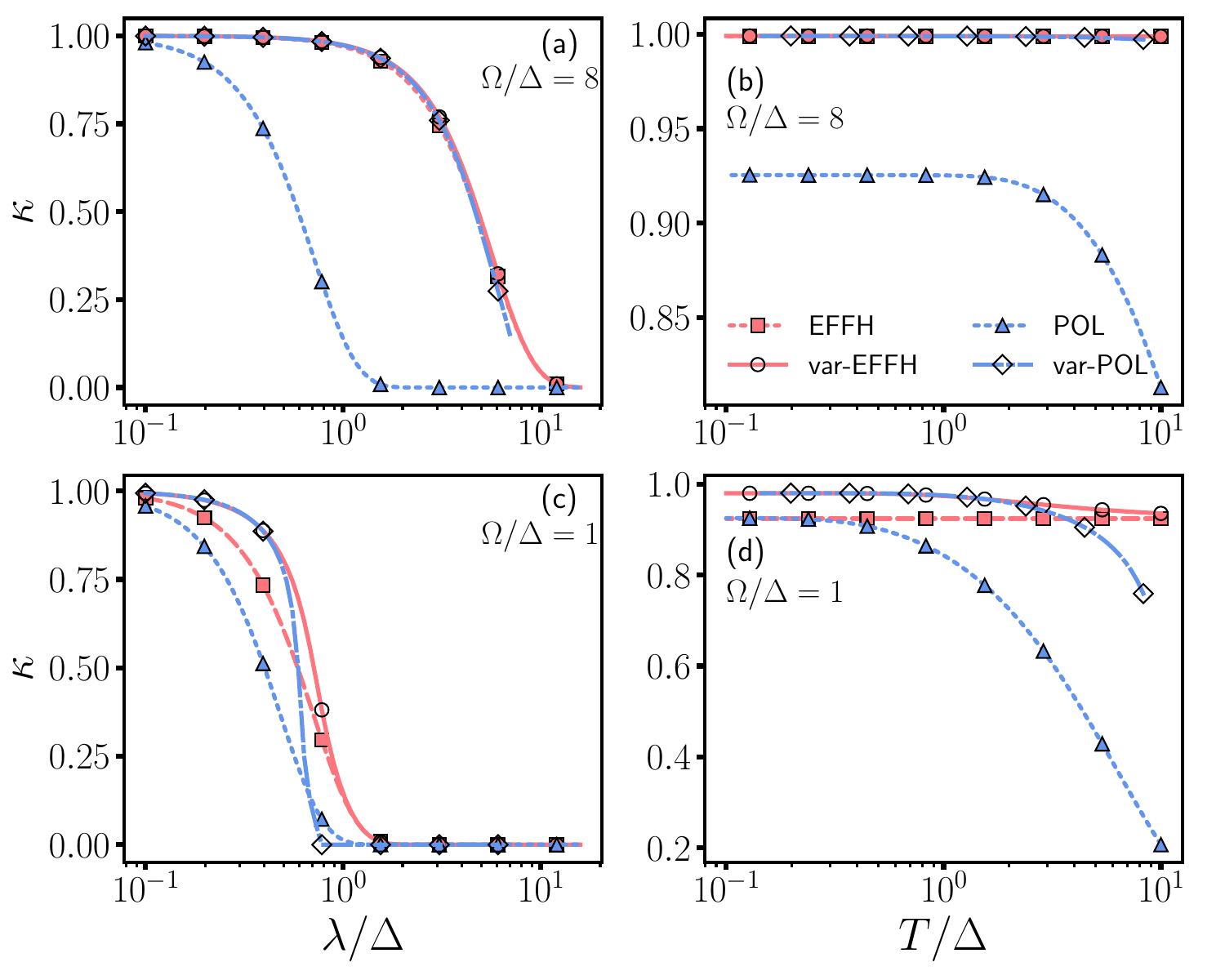}
\caption{The renormalization factor $\kappa$ for the generalized spin-boson model computed via Eq.~\eqref{eq:kappa} for the polaron (blue triangles) and variational polaron (diamonds). For the effective approach (red squares) and the variational effective approach (circles) we employ Eq.~\eqref{eq:kappaEff}. $\kappa$ is plotted as a function of (a)-(c) coupling strength $\lambda$ and (b)-(d) temperature $T$. Results are shown for $\Delta = 1$, $\theta = \pi/4$, $T = \Delta$ (a)-(c), and $\lambda = 0.2\Delta$ (b)-(d). 
}
\label{fig:fig2}
\end{figure}

%=========================
\begin{figure*}
    \centering
\includegraphics[width=1.8\columnwidth]{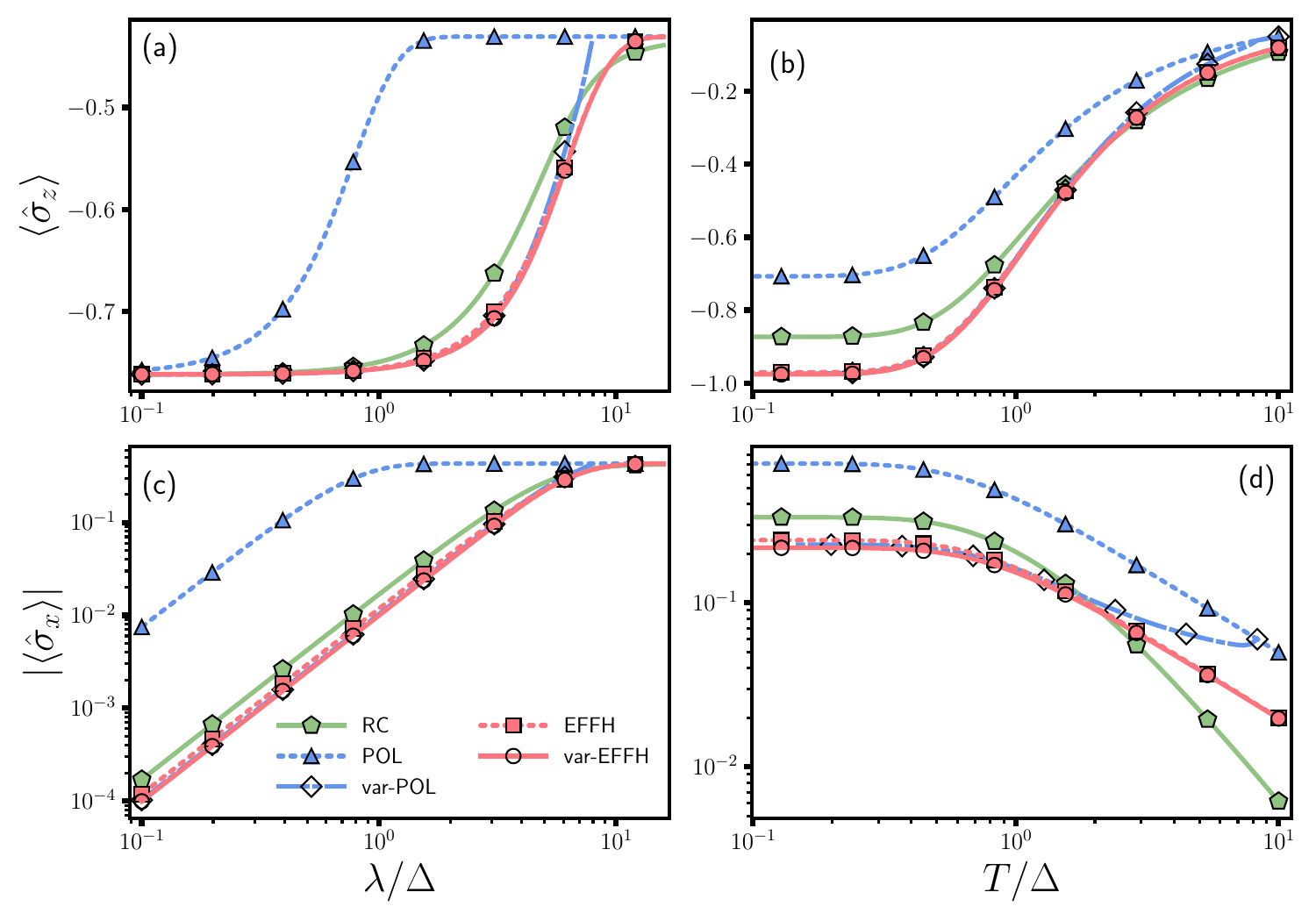}
\caption{The $\langle \hat{\sigma}_x \rangle$ and $\langle \hat{\sigma}_z \rangle$ spin polarizations for the generalized spin-boson model as a function of (a) and (c) coupling strength $\lambda$ and (b) and (d) temperature $T$. Results are shown for the effective approach (red squares), variational effective approach (circles), polaron (blue triangles) and variational polaron (diamonds) and the reaction coordinate method (green pentagons). Parameters are: $\Delta = 1$, $\theta = \pi/4$, %$\gamma = 0.0075$ 
$\Omega = 8$, $T = 1$ (a-c), and $\lambda = 4$ (b-d).
%\textcolor{orange} {...abs value for $\sigma_x$.}
%\textcolor{orange}{
%In all panels we need to use
%EFFH, var-EFFH, POL and var-POL.}
}
\label{fig:fig3}
\end{figure*}

%Show something to imply that at lower Omega the var-effH > effH. 
%Plot sigz, sigx as a function of omega?

\begin{figure*}
    \centering
\includegraphics[width=1.8\columnwidth]{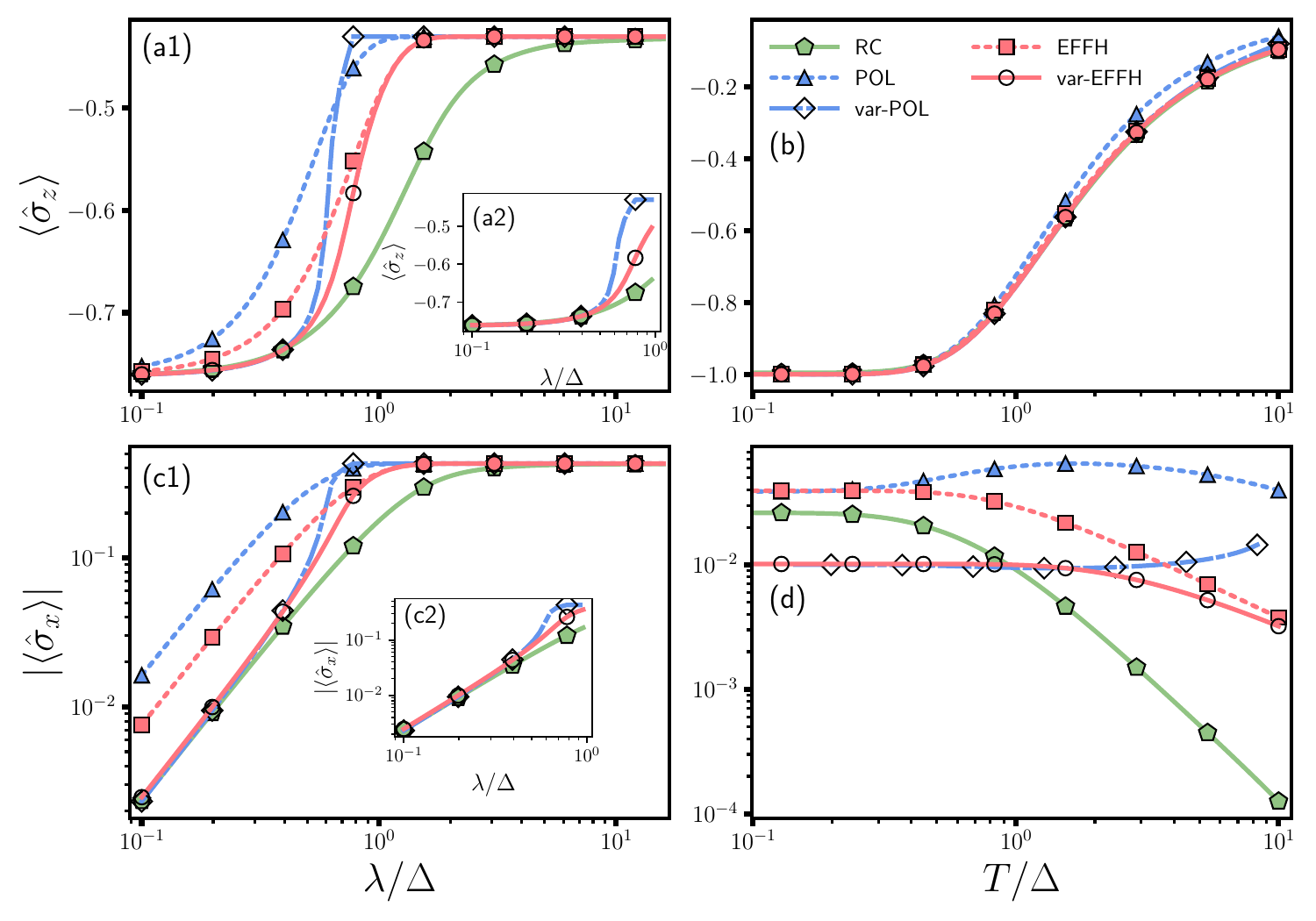}
\caption{The $\langle \hat{\sigma}_x \rangle$ and $\langle \hat{\sigma}_z \rangle$ spin polarizations for the generalized spin-boson model as a function of (a) and (c) coupling strength $\lambda$ and (b) and (d) temperature $T$. Results are shown for the effective approach (red squares), variational effective approach (circles), polaron (blue triangles) and variational polaron (diamonds) and the reaction coordinate method (green pentagons). Parameters are: $\Delta = 1$, $\theta = \pi/4$, %$\gamma = 0.0075$ 
$\Omega = 1$, $T = 1$ (a-c), and $\lambda = 0.2$ (b-d).
The small panels (a2) and (c2) zoom over the weak-coupling regime.
%\textcolor{orange} {...abs value for $\sigma_x$}
%In panels (a) and (c), pls put a box in the low lambda results to highlight the weak-intermediate regime where the var-EFFH is successful. You may be able to add it directly on the Pdf. 
%+ In all panels we need to use
%EFFH, var-EFFH, POL and var-POL.}
%{Use a dotted line for EFF, similarly to pol. Can you also please send me the results for $\Omega=2$?}
}
\label{fig:fig4}
\end{figure*}
%==================================

\section{MFGS: Analytical results}
\label{sec:anal}

We found that for the generalized spin-boson model, the EFFH 
and the POL methods generate the same form for the Hamiltonian of the system, Eq.~\eqref{eq:HSBVeff} and Eq.~\eqref{eq:HSP}, respectively. The only difference between them is the scheme for achieving the parameter $\kappa=\{\kappa_E,\kappa_P\}$, and the resulting value.
We recall that $\kappa$ embodies the impact of the system-bath coupling on the system through a renormalization effect to level splitting and the generation of new terms.
Using the EFFH/POL form for the system's Hamiltonian, we calculate the MFGS, arriving at~\cite{PRXRC}
\bea
    \hat\rho_{S} = \frac{1}{2}\left[ \hat{I} - \frac{(\vec{v}\cdot\vec{\sigma})}{\vert \vec{v} \vert}\tanh\left(\frac{\beta\Delta}{2} \vert \vec{v} \vert\right)\right],
\eea
where   $\vec{\sigma} = (\hat{\sigma}_x,\hat{\sigma}_y,\hat{\sigma}_z)$ and 
\bea
\nonumber
\vec{v} = [(1-\kappa)\sin(2\theta),0,(1+\kappa) + (1-\kappa)\cos(2\theta)].
\\
\eea
Here, $\hat{v}$ is the unit vector associated to $\vec{v}$ of magnitude 
\bea
\vert \vec{v} \vert = 
\sqrt{2(1+\kappa^2) + 2(1-\kappa^2)\cos(2\theta)}.
\eea
Using these expressions and choosing as an example the angle $\theta=\pi/4$, we get a closed-form expression for the polarization, covering its behavior at arbitrary coupling strength, 
\bea
\langle \hat \sigma_z\rangle = -\frac{1}{\vert \vec{v} \vert} (1+\kappa) \tanh\left(\frac{\beta \Delta}{2}\vert \vec{v} \vert\right).
\label{eq:pol}
\eea
This result approaches the canonical solution 
in the ultraweak coupling limit, which corresponds to $\kappa=1$,
\bea
\langle \hat \sigma_z\rangle \xrightarrow{\kappa\to 1} -\tanh\left(\beta \Delta\right).
\eea
In contrast, in the ultrastrong coupling limit we get
\bea
\langle \hat \sigma_z\rangle \xrightarrow{\kappa\to 0} -\frac{1}{\sqrt 2}  \tanh\left(\frac{\beta \Delta\sqrt 2}{2}\right),
\eea
which approaches (in magnitude) $1/\sqrt{2}$ at low temperatures and $\beta\Delta/2$ at high temperatures.

As for coherences, we find that (for the particular case of $\theta = \pi/4$)
\bea
\langle \hat \sigma_x\rangle = -\frac{1}{\vert \vec{v} \vert} (1-\kappa) \tanh\left(\frac{\beta \Delta}{2}\vert \vec{v} \vert\right).
\label{eq:coh}
\eea
While in the weak coupling limit the coherences vanish, in the ultrastrong limit we find that they sustain the same value as that of the polarization,
\bea
\langle \hat \sigma_x\rangle \xrightarrow{\kappa\to 0} -\frac{1}{\sqrt 2}  \tanh\left(\frac{\beta \Delta\sqrt 2}{2}\right).
\eea
These expressions were derived for $\theta = \pi/4$; the behavior of the expectation values for any $\theta$ is straightforward to obtain~\cite{PRXRC}.

At the mathematical level, the expressions for the polarization and the coherence are identical between the POL and the EFFH methods, both variational and non-variational. However, at finite coupling we have that $0<\kappa<1$ and so the actual values of $\kappa_E$ and $\kappa_P$
differ, leading to quantitative deviations between the methods.

%=================================================================
\section{Simulations}
\label{sec:simul}

We established in Fig.~\ref{fig:HEOM} the validity of the numerical reaction coordinate approach \cite{NickRC} on the equilibrium spin-boson model by benchmarking it against numerically-exact simulations. In this Section, our focus is on the var-EFFH method, and we ask three questions:
(i) How well does the var-EFFH method perform compared to RC simulations? 
We assume that the latter results are very close to numerically-exact simulations in our parameter range.
(ii) What is the range of parameters where it is imperative to employ the var-EFFH tool, compared to the non-variational treatment? 
(iii) How do the var-EFFH and the var-POL methods compare in their ability to treat strong coupling effects? 
%Again, benchmarked against RC simulations.

We begin in Fig.~\ref{fig:fig2} by comparing parameters in the effective models generated by the var-POL and the var-EFFH methods, as well as with their non-variational analogs.
We present simulations of the renormalization parameters $\kappa_P$ and $\kappa_E$, defined respectively in Eq.~\eqref{eq:kappa} for the polaron method and in  Eq.~\eqref{eq:kappaEff} for the EFFH treatment.
In the spin-boson model, this parameter uniquely impacts the original Hamiltonian by system-bath coupling effects.
The renormalization parameters are studied as a function of the coupling strength in Fig.~\ref{fig:fig2}(a) and Fig.~\ref{fig:fig2}(c) and as a function of temperature in Fig.~\ref{fig:fig2}(b) and Fig.~\ref{fig:fig2}(d), for the particular choice of $\theta = \pi/4$. We show both the ``bare" parameters without a variational treatment and variational results.  We test the behavior of $\kappa$ for both large $\Omega$ in  Fig.~\ref{fig:fig2}(a)-(b) and small $\Omega$ Fig.~\ref{fig:fig2}(c)-(d).

Beginning with the $\lambda$ dependence, in Fig.~\ref{fig:fig2}(a), we report a nearly-perfect agreement between the EFFH and var-EFFH for  $\Omega \gg T,\Delta$, suggesting that the variational scheme is not necessary in this regime, which is expected.
Furthermore, the var-POL approach  predicts very similar trends for the coupling dependence, from ultraweak well into the intermediate coupling regime. However in the ultrastrong coupling limit the var-POL algorithm 
did not converge.
%{\textcolor{red}{
% the graph does not show the previously mentioned ''collapse" to the normal polaron}}
%collapses to that of the full polaron approach, which results in a different approach to the ultrastrong coupling limit.
%
Lastly, the full polaron  displays stark numerical deviations. This is to be expected since a full-polaron transform is suitable for Ohmic spectral functions \cite{Legget,Weiss} but not for (narrow) Brownian spectral functions. 
% DD Need to talk about it more
Repeating the analysis for small $\Omega$, of order of both the spin splitting energy and the temperature, we find that the var-EFFH brings a smaller renormalization ($\kappa \approx 1$) than the full EFFH, with more qualitative deviations of the var-POL method at strong coupling.

In Fig.~\ref{fig:fig2}(b) and Fig.~\ref{fig:fig2}(d) we display the temperature dependence of $\kappa$. We point out the lack of temperature dependence in the non-variational EFFH approach;  the influence of the bath's temperature does not come into the protocol of generating the effective Hamiltonian in Ref. \cite{PRXRC}. In contrast, the var-EFFH method that we introduced in this work shows temperature dependence due to the inclusion of the temperature and other system parameters in the transcendental equation,  Eq.~\eqref{eq:kappa}. 
The dependence of $\kappa_E$ on temperature is more pronounced at high temperatures and less so at low temperatures, which is expected.
At the chosen coupling strength, the dependence on temperature is modest.
%
%This suggests that there is a temperature component to the EFFH that we are able to capture with our extension, thus providing a more accurate prediction to strong coupling physics by incorporating an additional temperature dependence into the renormalization parameters. 
% 
The var-POL approach also displays a temperature dependence: At low temperature, the induced renormalization effects are similar to those observed by the var-EFFH method.  Deviations show up at higher temperatures ($T/\Delta > 1$) and for small $\Omega$, where there is a rapid suppression in the var-POL renormalization parameter.  
%NAS need to explain or talk about the differences at high T. Why is it suppressing so much?
Lastly, the full polaron transformation significantly overestimates the effect of dressing, especially at high temperatures. 

%does not predict a temperature dependence and performs quite differently compared to the other three approaches, which we may establish as a symptom of selecting a Brownian bath.
%We conclude from Fig.~\ref{fig:fig2} that the var-EFFH performs similarly to the EFFH but contains an extra dependence on additional parameters in the model, such as the temperature.
%
%NAS Should the full-polaron be included since we are using a Brownian bath and the full polaron does not do well with ohmic-type reservoirs? It seems like not a fair comparison. The var-polaron is ok though 

To study the MFGS, we analyze physical observables of the spin, $\langle \hat{\sigma}_z \rangle$ and $\langle \hat{\sigma}_x \rangle$,
displayed in Fig.~\ref{fig:fig3} and Fig.~\ref{fig:fig4} for high and low characteristic bath frequencies, respectively. 
  % 
 %So far we have been discussing parameters relating to the respective methods, and similarities between the polaron and EFFH techniques. Now, we would like to see how they fair at predicting physical observables. 
We compare the following approaches: The non-variational EFFH and POL methods, along with their variational analogs. We further display RC simulations (green pentagons) serving as a benchmark to verify the numerical accuracy of each approach. 
In Fig.~\ref{fig:fig3}(a) and Fig.~\ref{fig:fig3}(c) we display the coupling-strength dependence of the spin polarization and coherences. We  report a near-perfect agreement between the EFFH, var-EFFH and the var-POL methods, which are also qualitatively similar to the RC technique. On the other hand, the full polaron method falls short again, underestimating the spin polarization and overshooting the coherence. 
The temperature dependence of the spin is presented in Fig.~\ref{fig:fig3}(b) and Fig.~\ref{fig:fig3}(d). In Fig.~\ref{fig:fig3}(b) we see similar trends as before, namely, the spin polarization is well approximated by the EFFH, var-EFFH and the var-POL approach, with relatively small deviations from the RC method; the full polaron approach shows once again larger discrepancies. As for the coherences shown in Fig.~\ref{fig:fig3}(d), they are not well captured in the high temperature limit by any of the approximate approaches: The  scaling of the coherences with temperature at high temperature, predicted by the var-EFFH method is different from that of the RC technique. 
This deviation can be rationalized by recalling that the EFFH method is not suitable to high temperature settings given the truncation of the reaction coordinate to its ground state. %This reasoning falls in line with the fact that the EFFH method can reasonably capture both the polarization and coherence at low temperatures. 
%The ultraweak coupling limit predicts zero coherences, thus the ability of the EFFH method to capture those terms that are purely related to strong coupling effect is worth noting.

Next, in Fig.~\ref{fig:fig4} we study analogous results compared to Fig.~\ref{fig:fig3}, yet for small bath frequency, with $\Omega$ being comparable to both $\Delta$ and $T$.
The EFFH method is not expected to perform well in this parameter range given that it is based on a complete truncation of the reaction coordinate. This truncation is justified when there is a separation of energy scales of the bath from the system with high levels of the reaction coordinate (above thermal energy) approximately not populated.
Focusing first on the behavior of polarization and coherences as a function of coupling strengths in Fig.~\ref{fig:fig4}(a) and Fig.~\ref{fig:fig4}(c), we note on the substantial success of the var-EFFH to capture the weak-to-intermediate coupling regime. The var-POL method is also successful in this regime, but at a certain value of the coupling it ``collapses" into the inaccurate non-variational polaron result. %Furthermore, the predicted results are significantly less accurate than any of the two the EFFH methods.
As for the temperature dependence shown in Fig.~\ref{fig:fig4}, we focus on the weak coupling limit where the methods perform relatively well. We note that the polarization in Fig.~\ref{fig:fig4}(b) is very well captured by all the approaches employed. However, coherences in Fig.~\ref{fig:fig4}(d) are captured only qualitatively by the variational and non-variational EFFH methods. Nevertheless, this moderate success should be contrasted with the fact that the POL and the var-POL methods are {\it unreliable} in the small $\Omega$ case beyond low temperatures, providing incorrect (non-monotonic) trends.    

Returning to questions posed at the beginning of this Section, our results point towards the following conclusions:
(i) The var-EFFH method can be used with confidence across the entire system-bath coupling range as long as $\Omega>T, \Delta$.  Furthermore, it can be employed at  smaller values of $\Omega$ provided one is interested only in the regime of weak-to-intermediate coupling energy.
(ii) The non-variational EFFH method is powerful and qualitatively correct whenever $\Omega \gg T$. Given its analytical form and simplicity, we argue in favor of this approach over the var-EFFH technique, since the latter  requires a numerical optimization procedure to build its parameter. 
(iii) The non-variational polaron approach should be avoided if one is interested in the description of equilibrium states at strong coupling when using structured spectal functions. The var-POL technique in contrast performs well over a broad range of both coupling strength and temperature. Its main caveat lies in its  ``collapse"  to full polaron results (or altogether non-convergence) in difficult parameter regimes, e.g., at ultrastrong couplings or at high temperatures, unlike the EFFH method that is always well behaved.

%We conclude by noting that even though the EFFH does not contain any dependence on $T$ and other model parameters, it still correctly predicts, at least approximately, the correct physics from ultraweak to ultrastrong with such a minimal model. While the inclusion of additional parameters using var-effH gives additional analytical intuition into the problem (See Eq. (\ref{eq:C2})) and performs similarly well in the regimes where the effH is appropriate. 
%=============================

%============================
\section{Discussion and Summary}
\label{sec:Summ}

We have focused on approximation schemes for the equilibrium state of a quantum system, the so-called MFGS. 
Our main focus has been the EFFH method that was first introduced in Ref.~\cite{PRXRC}, which is semi-analytical and handles strong system-bath coupling effects.
Using the spin-boson model as a case study, we validated the EFFH method from weak to strong coupling, 
developed a variational flavor that extends the regime of applicability of the original method and compared it to the well-known polaron-transform treatment. 
In more details, our contributions are threefold:

(i) First, we examined the effectiveness of the EFFH method in building an approximate MFGS of the general spin-boson model by benchmarking it against the numerically-exact HEOM. In addition to the excellent quantitative agreement in the ultraweak and ultrastrong system-bath coupling limits, we observed a correct qualitative behavior at all intermediate couplings. Furthermore, we emphasize that for the specific temperatures in the range $\Delta\approx T\ll \Omega$ 
%and when $\lambda\ll\Omega$, % DD2 no problem even when lambda --> infty (>> Omega)
we find quantitatively-accurate results of the EFFH MFGS in all system-bath coupling strength. 

(ii) The basic drawback of the EFFH treatment as introduced in Ref.~\cite{PRXRC} is that it is limited to parameter regimes where the typical energy scale of the system ($\Delta$) is close to the bath temperature ($T$) while sitting below  the characteristic frequency of the bath ($\Omega$). To relax this limitation, we developed here the var-EFFH method where the original, full polaron transform, enacted after the RC mapping, is replaced with a variational one. The consequence of the variational polaron procedure is the dressing of the original system parameters with not only the residual coupling ($\lambda$) and the characteristic frequency of the bath ($\Omega$), but also with the bath temperature ($T$). As a result, we extended the applicability of the original EFFH method even to parameter regimes where $\Omega\simeq \Delta$. In particular, the var-EFFH method captures the MFGS in the weak-to-intermediate coupling regime especially well. However, although the var-EFFH method can be used even when $\Omega$ is not large relative to the system energy scale $\Delta$ and the bath temperature $T$, the chosen values for $\Delta$ and $T$ must still be on the same scale, particularly when studying coherences.
High temperature effects cannot be well captured given the extreme truncation of the reaction coordinate, while low-temperature effects are still missing in the construction of the MFGS; possibly due to the development of strong system-bath correlations. These aspects require further investigation. 

(iii) Equipped with the new var-EFFH technique, we compare its efficacy in describing the MFGS against the popular polaron transformation. Both the POL/var-POL approach lead to the same system Hamiltonian structure as the EFFH/var-EFFH framework, where the only distinction lies in the value of the renormalized system parameters captured by $\kappa_P$ and $\kappa_E$, respectively. Qualitatively analogous simulation results were obtained in predicting the MFGS using the two methods. However, for structured spectral functions, the EFFH is advantageous over the polaron technique in accuracy, computational cost and stability. 
While the methods display the same physics, the EFFH is simpler and more robust in computations. 
 
We focused in this paper on the spin-boson model to understand and test the validity of the variational and non-variational EFFH and POL approaches. For the spin-boson model, we showed that these approaches built an identical system Hamiltonian, only with different parameters.  
In Appendix~\ref{app:4} and Appendix~\ref{app:5}, we further examine the Caldeira-Leggett model of a single harmonic mode bilinearly coupled to harmonic bath, and derive the system's Hamiltonian under the EFFH and the POL methods. 
% DD2 add reference to the CL model
We find that the two methods again reach the same result: The system's Hamiltonian after the EFFH mapping is identical to the original one only with a constant shift, thus indicating that the state of the system is still the standard Gibbs state, even at strong coupling.
This conclusion indeed holds at  high temperatures, $\Delta\ll T$, as discussed in Refs.~\cite{Lutz:2011Eq,HO1,HO2,HO3,HO4}.

% May add:
%We may use the classification of  ultraweak coupling, weak-intermediate, intermediate-strong and ultrastrong
%where the EFFH can treat the ultraweak, weak-intermediate and ultrastrong
%while still struggling with the intermediate-strong.
% energy scale lambda -->0, lambda --> infty and in between 
%lambda smaller to prder of Delta (weak to intermediate) and lambda order of Omega; intermediate to strong coupling.

% other spectral functions
A natural extension of our work is to test the validity of the var-EFFH treatment in obtaining the MFGS for a variety of bath spectral functions. In our work, we only studied Brownian-structured spectral functions where the RC is most-easily computed. To capture other situations, e.g., a  structured bath with multiple peaks or the popular Ohmic-type spectral functions, the procedure of the RC extraction has to be made into a numerical process. Once a reliable machinery of extracting RC modes from the bath is established, the EFFH method could be 
%var-POL transform of each RC mode should be % DD2 not very clear
trivially executed. 
% interacting components

Another avenue for the var-EFFH treatment is to study systems with multiple building blocks with or without direct interactions between them \cite{Archak,Kawai_2019}. In the case of noninteracting spins coupled to a common bath, bath-mediated spin-spin interaction terms arise beyond weak coupling. This was recently demonstrated {\it numerically}, using the RC method in Ref. \cite{Brenes_2023}, with applications to quantum thermometry. However, studying the development of cooperative effects with analytical tools such as the var-EFFH method is still missing. 
%
% DD2 EFFH of course holds beyond equilibrium, so this discussion will make people think that it does not.
%In addition, the EFFH method and it
%our prescription of more accurately 
%imprinting strong system-bath coupling effects onto the system could be applied to a nonequilibrium setting where a well-defined temperature does not exists. 
%A recent study indeed discussed generalizations of the classical definition of temperature to nonequilibrium quantum systems where a meaningful definition of temperature could be extracted \cite{Alipour_2021}. 

% noneq % DD2, replacing the above
The EFFH method can be readily used to treat systems out-of-equilibrium, as demonstrated in Ref. \cite{PRXRC}. With it generating closed-form expressions for the Hamiltonian with built-in strong system-bath couplings, it can be used to e.g., test concepts in nonequilibrium thermodynamics such as the identification of temperature-like measures in nonequilibrium quantum systems \cite{Alipour_2021}. 
%
% Dynamics
Another interesting application of the var-EFFH method is to focus on time-evolution problems, e.g., related to cavity-modified chemical reactivity \cite{ReichmanP23}. Recall that the method retains the original system-bath Hamiltonian structure [Eq.~\eqref{eq:SBeff}]. 
On the other hand, the usual polaron approach altered the system-bath coupling term as shown in Eq.~\eqref{eq:HSBPl}. Since the system-bath operators are distinct between the var-EFFH and the var-POL, a careful comparison to numerically-exact methods is required to establish the correct physics. 

%=============================================================
\begin{acknowledgements}
We acknowledge fruitful discussions with Anton Trushechkin, Bijay K. Agarwalla and Kartiek Agarwal. DS acknowledges support from an NSERC Discovery Grant and the Canada Research Chair program. The work of NAS and BM was supported by Ontario Graduate Scholarship (OGS). The work of MB has been supported by the Centre for Quantum Information and Quantum Control (CQIQC) at the University of Toronto. Computations were performed on the Niagara supercomputer at the SciNet HPC Consortium. SciNet is funded by: the Canada Foundation for Innovation; the Government of Ontario; Ontario Research Fund - Research Excellence; and the University of Toronto. 
\end{acknowledgements}
%======================

%=======================================

% Appendix A
\appendix
\section{The Hierarchical Equations of Motion}
\label{app:1}

The Hierarchical equations of motion (HEOM) is a numerically-exact technique used to solve for the dynamics and equilibrium state of open quantum systems. This method goes beyond standard quantum master equation approaches by avoiding the Born-Markov approximation, thus providing a rigorous and reliable tool to benchmark new methodologies in open quantum systems \cite{TanimuraHEOM,Kawai2020,TanimuraBook,Lambert2023}. 
The HEOM starts by discretizing the continuum of states in the environment and organizing them in a hierarchy of equations of motion for auxiliary density matrices to be solved simultaneously. Care must be taken to ensure a sufficient number of hierarchies are maintained in the dynamics to allow for the discretized environment to correctly model the reservoir. Beyond this, one assumes that the bath correlation functions are represented as a sum of exponential terms. The exact form of these exponential terms can be determined analytically for certain classes of spectral density functions. A sufficient number of exponential terms should be incorporated in the sum to ensure physical and converged results for the state of the system.
More concretely, following the style of Ref. \cite{Lambert2023}, the bath correlation function is given by
\bea
    C(t) &=& \langle \hat{X}(\tau+t) \hat{X}(\tau) \rangle_B
    \\ \nonumber
    &=& \int_0^{\infty} {\rm{d}}\omega J(\omega)\left[ \coth \left(\frac{\beta\omega}{2} \right) \cos(\omega t) - \ii\sin(\omega t) \right ],
\eea
% DD2
where $\hat X= \sum_{k} t_k (\hat c_k^{\dagger}+\hat c_k)$ according to Eq. \eqref{eq:h_total_model}.
Assuming that the real and imaginary parts of the bath correlation functions can be decomposed into a sum of exponential terms, we may write the above expression as
\bea
    C(t) = \sum_{k=1}^{N_R} c_k^R e^{-\gamma_k^R t} + \ii\sum_{k=1}^{N_I} c_k^I e^{-\gamma_k^I t},
\eea
where in the above expression, $N_R$ and $N_I$ are convergence parameters controlling the number of Matsubara terms in the real and imaginary parts of the expansion. Moreover, the expansion coefficients $c_k^R$ and $c_k^I$ as well as the Matsubara frequencies $\gamma_k^R$ and $\gamma_k^I$ can be both real or complex. 
For the Brownian bath model studied in the main text, these parameters are given by Eqs.~(24)-(27) in Ref.~\cite{Lambert2023}.

Using this, the $n$-th equation in the hierarchy can be constructed as 
\bea
    \dot{\rho}^n(t) &=& \left( -\ii\hat{H}_S^{\times} - \sum_{j = R,I} \sum_{k=1}^{N_j} n_{jk} \gamma_k^j \right)\rho^n(t) 
    \\ \nonumber
    &-& \ii\sum_{k=1}^{N_R} c_k^Rn_{Rk}\hat{S}^{\times} \rho^{n^-_{Rk}}(t) + \sum_{k=1}^{N_I} c_k^In_{Ik}\hat{S}^{\circ} \rho^{n^-_{Ik}}(t)
    \\ \nonumber
    &-& \ii\sum_{j = R,I} \sum_{k=1}^{N_j} \hat{S}^{\times}\rho^{n^+_{jk}}(t).
    \label{Eq:HEOM}
\eea
In the above expression, we use the following notation for the operators $\hat{O}^{\times}\bullet = [\hat{O},\bullet]$ and $\hat{O}^{\circ}\bullet = \{\hat{O},\bullet\}$. Furthermore, $n = (n_{R1},n_{R2},...,n_{RN},n_{I1},n_{I2},...n_{IN})$ is a multidimensional index used to label the auxiliary density matrices with each $n_{jk}$ taking values in the set $\{0,1,2,\cdots,N_c\}$, with $N_c$ being a convergence parameter indicating the number of hierarchies to include. Worthy of note, the state labelled by (0,...,0) corresponds to the system density matrix of interest. Furthermore, terms such as $\rho^{n^{\pm}_{jk}}(t)$ correspond to an auxiliary density matrix with index $n_{jk}$ raised or lowered by one.

We use the HEOM implementation from the Quantum Toolbox in Python (QuTiP) package \cite{Lambert2023, QuTip1, QuTip2} to solve for the steady state polarization in the generalized spin-boson model in order to benchmark the EFFH approach in Fig.~\ref{fig:HEOM}.

%======================================================

\section{Second order correction to the Gibbs state}
\label{app:2}

In Fig.~\ref{fig:HEOM} we compared the second order correction to the mean force Gibbs state against our effective Hamiltonian approach. The correction to the MFGS was derived in Ref.~\cite{JanetPRL}, and we briefly summarize it here.

For a given system Hamiltonian $\hat{H}_S$ and coupling operator $\hat{S}$, the second order correction to the mean force Gibbs state is given by
\bea
    \nonumber
    \hat \rho^{(2)}_S &=& \hat \tau_S + \epsilon^2 \beta \sum_n \hat{\tau}_S (\hat{S}_n\hat{S}_n^{\dagger} - \Tr_S[\hat{\tau}_S \hat{S}_n\hat{S}_n^{\dagger}])D_{\beta}(\omega_n) 
    \\ 
    &+& \epsilon^2\sum_n [\hat{S}_n^{\dagger},\hat{\tau}_S \hat{S}_n] \frac{dD_{\beta}(\omega_n)}{d\omega_n}
    \\ \nonumber
    &+& \sum_{n \neq m} ([\hat{S}_m, \hat{S}_n^{\dagger} \hat{\tau}_S] + {\rm h.c.})\frac{D_{\beta}(\omega_n)}{\omega_{nm}}.
\eea
%Should we show to bound illustrating when the expansion is valid?
This result is obtained by performing a perturbative expansion of the mean force Gibbs state to second order in the system-bath coupling parameter $\epsilon$,
$  \hat{\rho}_S = \frac{\Tr_B\left(e^{-\beta\hat{H}}\right)}{Z}$, 
where $\hat{H}$ is the total Hamiltonian of the system and reservoir. In the above expression,
% DD2
$\hat S= \sum_n \hat S_n$.
The operators $\hat{S}_n$ are the energy eigenoperators obtained from solving the eigenoperator equation $[\hat{H}_S,\hat{S}_n] = \omega_n \hat{S}_n$, with $\omega_n$ the eigenvalues. We further define $\omega_{nm}\equiv \omega_n-\omega_m$. 
Moreover, $\hat{\tau}_S$ is the standard weak coupling Gibbs state with respect to $\hat{H}_S$, and lastly, $D_{\beta}(\omega_n)$ contains nontrivial temperature dependence. For the case where $\hat{S}^2 \propto \hat{I}$ (such as the case in our study), we have
\bea
    D_{\beta}(\omega_n) = \int_0^{\infty} {\rm d}\omega J(\omega) \frac{\omega_n\coth(\frac{1}{2}\beta\omega) + \omega}{\omega^2 - \omega_n^2}.
\eea
For the generalized spin-boson model studied in the main text,  $\hat{S} = \hat{\sigma}_z\cos(\theta) + \hat{\sigma}_x\sin(\theta)$, for which the associated eigenoperators and eigenvalues are given by
\bea
    \hat{S}_1& =& \sin(\theta)\hat{\sigma}_+\,\,\,\,\,\, \omega_1 = 2\Delta
    \\
    \hat{S}_0 &=& \cos(\theta)\hat{\sigma}_z \,\,\,\,\,\, \omega_0 = 0
    \\
    \hat{S}_{-1} &=& \sin(\theta)\hat{\sigma}_- \,\,\,\,\,\, \omega_{-1} = -2\Delta.
\eea
As a result, one can write the second order correction of the mean force Gibbs state as \cite{JanetPRL}
\bea
    \hat \rho_S^{(2)} = \hat{\tau}_S + \frac{\langle \hat{\sigma}_x \rangle}{2} \hat{\sigma}_x + \frac{\langle \hat{\sigma}_z \rangle - \langle \hat{\sigma}_z \rangle_0}{2} \hat{\sigma}_z,
\eea
with the coefficients 
\begin{widetext}
\bea
    \langle \hat{\sigma}_x \rangle = - \frac{\epsilon^2\sin(2\theta)}{\Delta} \left[ \tanh(\beta\Delta)\int_0^{\infty}{\rm d}\omega J(\omega) \coth\left(\beta\omega/2\right) \frac{2\Delta}{\omega^2 - 4\Delta^2} - \int_0^{\infty}{\rm d}\omega J(\omega) \frac{\omega}{\omega^2 - 4\Delta^2}  + \int_0^{\infty}{\rm d}\omega \frac{J(\omega)}{\omega} \right].
    \label{Eq: sigx}
\eea
\bea
    \langle \hat{\sigma}_z \rangle - \langle \hat{\sigma}_z \rangle_0 &=&  2 \epsilon^2\sin^2(\theta) \Bigg[ \tanh(\beta\Delta)\int_0^{\infty}{\rm d}\omega J(\omega) \coth\left( \beta \omega/2\right) \frac{\omega^2 + 4\Delta^2}{(\omega^2 - 4\Delta^2)^2} - \int_0^{\infty}{\rm d}\omega J(\omega) \frac{4\omega\Delta}{(\omega^2 - 4\Delta^2)^2}  
    \\ \nonumber
    &+& \frac{\beta}{2}{\rm{sech}}^2(\beta\Delta) \int_0^{\infty}{\rm d}\omega J(\omega) \coth\left( \beta \omega/2\right) \frac{2\Delta}{\omega^2 - 4\Delta^2} \Bigg].
    \label{Eq: sigz}
\eea
%\end{widetext}
Note the difference in notation between these expressions and Ref.~\cite{JanetPRL}. Namely, the sign difference in the parameter $\theta$ and the factor of two in the spin-splitting. %However, given these two aspects the equations are consistent. 
In the main text, we computed the spin polarization and coherences, %along the $z$ and $x$ directions, 
which for this second order approach are represented by the solid black lines in Fig.~\ref{fig:HEOM}.
%============================================

%=====================================
% Appendix C
%\begin{widetext}
\section{Details on the self-consistent calculation in the var-POL MFGS}
\label{app:3}

We provide here  details on the var-POL method,
adding the 
derivation of the self-consistent equation used to solve for the optimal parameters $\{f_k\}$, Eqs.~\eqref{eq:tf}-\eqref{eq: F pi half and fourth}.
Recall that we minimize the Gibbs-Bogoliubov-Feynman upper bound on the free energy, which is given by 
\begin{equation}
    A_B = -\frac{1}{\beta}\ln \Tr\left(e^{-\beta\hat{H}^\text{pol}_0}\right)+\langle \hat{H}^\text{pol}_I\rangle_{\hat{H}^\text{pol}_0},
\end{equation}
where $\hat{H}^\text{pol}_0 = E_0\hat{I}_2+\hat{H}^\text{pol}_S+\hat{H}^\text{pol}_B$. Since the bath Hamiltonian does not depend on $\{f_k\}$, we can simply minimize the system free energy. Furthermore, since $\langle \hat{H}^\text{pol}_I\rangle_{\hat{H}^\text{pol}_0}$ is zero by construction, we end up by minimizing
\begin{equation}
    A_B = -\frac{1}{\beta}\ln Z_S,
\end{equation}
where $Z_S$ is the partition function for $E_0\hat{I}_2+\hat{H}^\text{pol}_S$. 
It is given by 
\begin{equation}
    Z_S = 2e^{-\beta E_0}\cosh\left(\frac{\beta\Delta\sqrt{1+\kappa_P^2-(\kappa_P^2-1)\cos(2\theta)}}{\sqrt{2}}\right).
\end{equation}
This leads to the free energy expression 
\begin{equation}
    A_B=-\frac{1}{\beta}\ln\left[2e^{-\beta E_0}\cosh\left(\frac{\beta\Delta\sqrt{1+\kappa_P^2-(\kappa_P^2-1)\cos(2\theta)}}{\sqrt{2}}\right)\right].
\end{equation}
Minimizing it with respect to $f_k$, $\frac{\partial A_B}{\partial f_k}=0$, we obtain
\begin{equation}
    \frac{2(f_k-t_k)}{\nu_k}-\frac{\Delta}{\sqrt{2}}\tanh\left(\frac{\beta\Delta\sqrt{1+\kappa_P^2-(\kappa_P^2-1)\cos(2\theta)}}{\sqrt{2}}\right)\frac{(1-\cos(2\theta))\kappa_P^2}{\sqrt{1+\kappa_P^2-(\kappa_P^2-1)\cos(2\theta)}}\left[-4\coth\left(\frac{\beta\nu_k}{2}\right)\frac{f_k}{\nu^2_k}\right]=0.
\end{equation}
Replacing $f_k=t_kF(\omega_k)$, and solving for $F(\nu_k)$, we find
\begin{equation}
    F(\nu_k) = \left[1+\sqrt{2}\Delta\tanh\left(\frac{\beta\Delta\sqrt{1+\kappa_P^2-(\kappa_P^2-1)\cos(2\theta)}}{\sqrt{2}}\right)\coth\left(\frac{\beta\nu_k}{2}\right)\frac{(1-\cos(2\theta))\kappa_P^2}{\nu_k\sqrt{1+\kappa_P^2-(\kappa_P^2-1)\cos(2\theta)}}\right]^{-1}.
\end{equation}
This result applies for a general choice of $\theta$. Specifically, for $\theta=\pi/2$ and $\theta=\pi/4$ the above condition simplifies to the results in the main text, Eq. \eqref{eq: F pi half and fourth}. 

To summarize, our transformed Hamiltonians in these two limits are 
\begin{equation}
\begin{cases}
    \hat{H}^\text{pol}= E_0\hat{I}_2+\Delta\kappa_P \hat{\sigma}_z+\sum_k\nu_k\hat{c}^\dagger_k\hat{c}_k+\Vec{V}\cdot\Vec{\sigma} & \text{for $\theta=\pi/2$}\\
     \hat{H}^\text{pol}= E_0\hat{I}_2+\frac{\Delta}{2}\left(1+\kappa_P\right)\hat{\sigma}_z+\frac{\Delta}{2}\left(1-\kappa_P\right)\hat{\sigma}_x
  +\sum_k\nu_k\hat{c}^\dagger_k\hat{c}_k+\Vec{V}\cdot\Vec{\sigma} & \text{for $\theta=\pi/4$},
    \end{cases}
\end{equation}
where $\Vec{\sigma}=\left(\hat{\sigma}_x,\hat{\sigma}_y,\hat{\sigma}_z\right)$ and 
\begin{equation}
\begin{cases}
    \Vec{V}=\left[\sum_k(t_k-f_k)(\hat{c}^\dagger_k+\hat{c}_k),-\Delta \sin(\hat{B}),\Delta\left(\cos(\hat{B})-\kappa_P\right)\right] & \text{for $\theta=\pi/2$}\\
    \Vec{V}=\left[\frac{\sum_k(t_k-f_k)(\hat{c}^\dagger_k+\hat{c}_k)}{\sqrt{2}}-\frac{\Delta}{2}\left(\cos(\hat{B})-\kappa_P\right),-\frac{\Delta}{\sqrt{2}}\sin(\hat{B}),\frac{\sum_k(t_k-f_k)(\hat{c}^\dagger_k+\hat{c}_k)}{\sqrt{2}}+\frac{\Delta}{2}\left(\cos(\hat{B})-\kappa_P\right)\right] & \text{for $\theta=\pi/4$}.
    \end{cases}
\end{equation}
We emphasize that the main distinction between the var-EFFH mapping and the var-POL transformation is the generated system-bath interaction Hamiltonian.

%==========================================
\section{Effective Hamiltonian theory of the Calderia-Leggett model}
\label{app:4}

In this Appendix, we derive the effective Hamiltonian of the Calderia-Leggett model, describing in particular a dissipative quantum harmonic oscillator \cite{Legget,Weiss}.
The EFFH allows capturing strong system-bath coupling effects within the system's Hamiltonian, thus building the MFGS. In the main text we showed that the MFGS resulting from the EFFH largely deviates from the Gibbs state beyond ultraweak coupling.
In this Appendix we will find out what that means for a fully harmonic model.

We begin with a harmonic-oscillator (HO) model for the system's Hamiltonian ($\hbar=1$ and $m=1$), but as we point out below Eq. \eqref{eq:effHO}, our conclusions hold for a general potential $V(\hat x)$,
\bea
    \hat{H}_S = \frac{1}{2}\omega^2_0\hat{x}^2 + \frac{1}{2}\hat{p}^2.
    \label{eq:HO1}
\eea
The position operator of the HO couples to the environment, $\hat{S} = \hat{x}$. 
Using the approach delineated in Ref. \cite{PRXRC} the effective Hamiltonian of the system is constructed as
\bea
    \hat{H}_S^{\rm eff} &=& e^{-\frac{\lambda^2}{2\Omega^2}\hat{x}^2} \sum_{n=0}^{\infty} \frac{\lambda^{2n}}{\Omega^{2n}n!} \hat{x}^n \left(\frac{1}{2}\omega^2_0\hat{x}^2 + \frac{1}{2}\hat{p}^2\right) \hat{x}^n   e^{-\frac{\lambda^2}{2\Omega^2}\hat{x}^2}
    \nonumber\\ 
    &=& \frac{1}{2}\omega_0^2 \hat{x}^2 + \frac{1}{2} e^{-\frac{\lambda^2}{2\Omega^2}\hat{x}^2} \sum_{n=0}^{\infty} \frac{\lambda^{2n}}{\Omega^{2n}n!} \hat{x}^n \hat{p}^2 \hat{x}^n   e^{-\frac{\lambda^2}{2\Omega^2}\hat{x}^2}.
\label{eq:HO2}
\eea
Since the position operator commutes with powers of itself, the first term is unaffected by the system-reservoir coupling. 
All changes occur in the second term due to the non-commutation of $\hat{x}$ and $\hat{p}$. 
We first need to learn how to deal with commutators of $\hat{p}^2$ and functions of $\hat{x}$, $f(\hat{x})$.
Acting on a test function,
we note that $[\hat{p}, f(\hat{x})] = -\ii f'(\hat{x})$, and
\bea
[\hat{p}^2, f(\hat{x})] = \frac{2}{\ii} f'(\hat{x})\hat{p} - f''(\hat{x}).
\eea
We use this result to rearrange the second term in Eq. (\ref{eq:HO2}),
\bea
&&e^{-\frac{\lambda^2}{2\Omega^2}\hat{x}^2} \sum_{n=0}^{\infty} \frac{\lambda^{2n}}{\Omega^{2n}n!} \hat{x}^n \hat{p}^2 \hat{x}^n   e^{-\frac{\lambda^2}{2\Omega^2}\hat{x}^2} = e^{-\frac{\lambda^2}{2\Omega^2}\hat{x}^2} \sum_{n=0}^{\infty} \frac{\lambda^{2n}}{\Omega^{2n}n!} \hat{x}^n ([\hat{p}^2,\hat{x}^n] + \hat{x}^n\hat{p}^2)e^{-\frac{\lambda^2}{2\Omega^2}\hat{x}^2}
 \nonumber\\
%&=& e^{-\frac{\lambda^2}{2\Omega^2}\hat{x}^2} \sum_{n=0}^{\infty} \frac{\lambda^{2n}}{\Omega^{2n}n!} (\hat{x}^n [\hat{p}^2,\hat{x}^n] + \hat{x}^{2n}\hat{p}^2)e^{-\frac{\lambda^2}{2\Omega^2}\hat{x}^2}
%\\
%&=& e^{-\frac{\lambda^2}{2\Omega^2}\hat{x}^2} \sum_{n=0}^{\infty} \frac{\lambda^{2n}}{\Omega^{2n}n!} \hat{x}^n [\hat{p}^2,\hat{x}^n] e^{-\frac{\lambda^2}{2\Omega^2}\hat{x}^2} + e^{\frac{\lambda^2}{2\Omega^2}\hat{x}^2}\hat{p}^2e^{-\frac{\lambda^2}{2\Omega^2}\hat{x}^2}
%\\
&&= e^{-\frac{\lambda^2}{2\Omega^2}\hat{x}^2} \left( \frac{2}{\ii}\sum_{n=0}^{\infty} \frac{\lambda^{2n}}{\Omega^{2n}(n-1)!} \hat{x}^{2n-1} \hat{p} - \sum_{n=0}^{\infty} \frac{\lambda^{2n}}{\Omega^{2n}(n-2)!} \hat{x}^{2n-2} \right) e^{-\frac{\lambda^2}{2\Omega^2}\hat{x}^2}+e^{\frac{\lambda^2}{2\Omega^2}\hat{x}^2}\hat{p}^2e^{-\frac{\lambda^2}{2\Omega^2}\hat{x}^2}.
\eea
Next, we use properties of Poisson distributions 
to evaluate the two sums in the above expression,
\bea
    \sum_{n=0}^{\infty} \frac{\lambda^{2n}}{\Omega^{2n}(n-1)!} \hat{x}^{2n-1} &=& %\sum_{n=0}^{\infty} \frac{(\lambda^{2})^n}{(\Omega^{2})^n(n-1)!} (\hat{x^2})^n \hat{x}^{-1} \frac{n}{n}
    %\\
    %&=& \sum_{n=0}^{\infty} n (\frac{\hat{x}^2 \lambda^{2}}{\Omega^{2}})^n \frac{1}{n!} \hat{x}^{-1}
    %\\
    %&=& 
    \frac{\lambda^2}{\Omega^2} \hat{x} e^{\frac{\lambda^2}{\Omega^2}\hat{x}^2},
\eea
and
%Here we used properties of poisson distributions to sum the series.
\bea
    \sum_{n=0}^{\infty} \frac{\lambda^{2n}}{\Omega^{2n}(n-2)!} \hat{x}^{2n-2} &=& \frac{\lambda^4}{\Omega^4} \hat{x}^2 e^{\frac{\lambda^2}{\Omega^2}\hat{x}^2}.
\eea
%which is obtained from similar manipulations to the first sum.
Combining these expressions we obtain,
\bea
\hat{H}_S^{\rm eff} &=& \frac{1}{2}\omega^2_0 \hat{x}^2 - \ii\frac{\lambda^2}{\Omega^2} \hat{x} e^{\frac{\lambda^2}{2\Omega^2}\hat{x}^2}\hat{p}e^{-\frac{\lambda^2}{2\Omega^2}\hat{x}^2} - \frac{1}{2}\frac{\lambda^4}{\Omega^4}\hat{x}^2 + \frac{1}{2}e^{\frac{\lambda^2}{2\Omega^2}\hat{x}^2}\hat{p}^2e^{-\frac{\lambda^2}{2\Omega^2}\hat{x}^2}.
%\\
%&=& \frac{1}{2}(\omega_0 - \frac{\lambda^4}{\Omega^4}) \hat{x}^2 - i\frac{\lambda^2}{\Omega^2} \hat{x} e^{\frac{\lambda^2}{2\Omega^2}\hat{x}^2}\hat{p}e^{-\frac{\lambda^2}{2\Omega^2}\hat{x}^2} + \frac{1}{2}e^{\frac{\lambda^2}{2\Omega^2}\hat{x}^2}\hat{p}^2e^{-\frac{\lambda^2}{2\Omega^2}\hat{x}^2}
\eea
Lastly, we note that 
\bea
e^{\frac{\lambda^2}{2\Omega^2}\hat{x}^2}\hat{p}e^{-\frac{\lambda^2}{2\Omega^2}\hat{x}^2} &=& \hat{p} + \ii\frac{\lambda^2}{\Omega^2}\hat{x},
\eea
and
\bea
    e^{\frac{\lambda^2}{2\Omega^2}\hat{x}^2}\hat{p}^2e^{-\frac{\lambda^2}{2\Omega^2}\hat{x}^2} &=& \hat{p}^2 +2\ii\frac{\lambda^2}{\Omega^2} \hat{x}\hat{p} - \frac{\lambda^4}{\Omega^4}\hat{x}^2 + \frac{\lambda^2}{\Omega^2}.
    \label{eq:effHO}
\eea
Combining all of the above we get the final expression for the effective system Hamiltonian
\bea
    \hat{H}_S^{{\rm eff}} = \hat{H}_S + \frac{\lambda^2}{2\Omega^2}.
\eea
Notably, we re-obtained a HO system, only with a constant shift.
% DD We need to be more careful as to what the MFGS is. One needs to subtract the "reorganization" energy.
%This means that the MFGS predicted by the EFFH method is trivially given by 
%$\hat{\rho}_S=\frac{e^{-\beta \hat{H}_S^{\rm eff}}}{ {\rm Tr}_S \left(e^{-\beta \hat{H}_S^{\rm eff}}\right)}$, with a harmonic system, similarly to the weak-coupling case.
The MFGS is given by this effective Hamiltonian, along with term representing the ``reorganization energy". This equilibrium state will be the same as that obtained in the weak coupling limit.
The {\it dynamics} of the HO at strong coupling would show a more involved behavior:
Given $J(\omega)$ for the original model, Eq. \eqref{eq:HO1}, after the EFFH mapping we retrieve this Hamiltonian, yet with a {\it different} spectral function for the bath. Thus, the argument is that one can study the dynamics of the newly mapped problem using weak coupling tools to capture strong coupling effects in the original picture.

Two comments are in place:
(i) The EFFH mapping does not affect any Hamiltonian of $V(\hat x)$ form, as long as the interaction with the bath is done through $\hat x$.
Thus, the EFFH method does not provide any advantage into calculating strong coupling effects for this class of models. 
However, once more the dynamics as predicted by the EFFH method would capture strong coupling effects given that it now takes into account a modified spectral function. 
(ii) An analogy to the present case is the pure decoherence model, where again the EFFH does not impact the system's Hamiltonian 
[compare Eq. \eqref{eq: general spin boson} to \eqref{eq:SBS} at $\theta=0$]
yet working with different spectral functions (before and after the EFFH process) do lead to differences in the decoherence dynamics as demonstrated in Ref. \cite{RCM}. 
%============

% Appendix E
\section{Variational polaron transformation of the harmonic model}
\label{app:5}

We apply here the variational polaron approach as presented in Sec. \ref{sec:polaron} onto the harmonic-oscillator model, and show that the resulting system's Hamiltonian, thus the MFGS, agree with the outcome of the EFFH method, as described in Appendix \ref{app:4}.
It is convenient to work here in second-quantized form as in Eq. \eqref{eq:h_total_model}. The system (a single harmonic mode of frequency $\nu_0$) and its interaction with the bath are given by
\begin{equation}
    \hat{H}_S = \nu_0 \hat{c}^\dagger \hat{c}\quad\text{and}\quad \hat{S} = \frac{1}{\sqrt{2\nu_0}}\left(\hat{c}^\dagger+\hat{c} \right) = \ell\left(\hat{c}^\dagger+\hat{c}\right).
\end{equation}
We perform the var-POL transform with $U=\exp(-\ii\hat{S}\hat{B}/2)$ where $\hat{B}=2\ii\sum_k\frac{f_k}{\nu_k}\left(\hat{c}^\dagger_k-\hat{c}_k\right)$, on the following Hamiltonian,
\begin{equation}
\begin{aligned}
    \hat{H} =& \nu_0\hat{c}^\dagger \hat{c}+\sum_k \nu_k\left(\hat{c}^\dagger_k+\frac{t_k}{\nu_k}\hat{S}\right)\left(\hat{c}_k+\frac{t_k}{\nu_k}\hat{S}\right),
    \end{aligned}
\end{equation}
and obtain
\begin{equation}
    \hat{H}^\text{pol}=\nu_0\hat{c}^\dagger\hat{c}-\frac{\ii \ell}{2}\left(-\hat{c}^\dagger+\hat{c}\right)\hat{B}-\left(-\frac{\ii \ell}{2}\right)^2\hat{B}^2+\hat{S}^2\sum_k\frac{(t_k-f_k)^2}{\nu_k}+\sum_k \nu_k\hat{c}^\dagger_k\hat{c}_k+\hat{S}\sum_k(t_k-f_k)(\hat{c}^\dagger_k+\hat{c}_k).
\end{equation}
Considering the first four terms, we calculate their thermal average with respect to the thermal state of the bath and obtain
%over the bathfor the first three terms which corresponds to the transformed system Hamiltonian,  % DD2 four terms
% 
\begin{equation}
    \hat{H}^\text{pol}_S = \left[\nu_0+2\ell^2\sum_k\frac{(t_k-f_k)^2}{\nu_k}\right]\hat{c}^\dagger \hat{c}+\ell^2\sum_k\frac{(t_k-f_k)^2}{2}\left[\hat{c}^\dagger\hat{c}^\dagger +\hat{c}\hat{c}\right]+\ell^2\sum_k\left[\frac{(t_k-f_k)^2}{\nu_k}+\frac{f^2_k}{\nu^2_k}(1+2n(\nu_k))\right].
\end{equation}
Here, $n(\nu_k)=1/(e^{\beta\nu_k}-1)$ is the Bose-Einstein distribution. For the full polaron transform ($f_k=t_k$), this expression reduces to
\begin{equation}
\hat{H}^\text{pol}_S = \nu_0\hat{c}^\dagger \hat{c}+\ell^2\sum_k\frac{t^2_k}{\nu^2_k}[1+2n(\nu_k)].
\end{equation}
Hence, we end up with the original system Hamiltonian with a constant energy shift that depends on the system-bath coupling strength and the bath temperature.

It is significant to note that the var-POL approach generates a system Hamiltonian with additional terms, beyond the original harmonic Hamiltonian. However, to understand their impact one needs to first achieve the variational parameters $f_k$ through a self-consistent approach, which is not a trivial task.

%{\textcolor{red}{Write the expression under full polaron}}

\end{widetext}

%====================================

\end{document}